\documentclass[preprint2]{aastex}
\usepackage{graphicx}
\begin{document}

\title{AN AMMONIA SPECTRAL MAP OF THE L1495-B218 FILAMENTS IN THE TAURUS MOLECULAR CLOUD : I. PHYSICAL PROPERTIES OF FILAMENTS AND DENSE CORES}

\author{Young Min Seo\altaffilmark{1}, Yancy L. Shirley\altaffilmark{1,2}, Paul Goldsmith\altaffilmark{3}, Derek Ward-Thompson\altaffilmark{4}, Jason M. Kirk\altaffilmark{4}, Markus Schmalzl\altaffilmark{5}, Jeong-Eun Lee\altaffilmark{6}, Rachel Friesen\altaffilmark{7}, Glen Langston\altaffilmark{8}, Joe Masters\altaffilmark{9}, Robert W. Garwood\altaffilmark{9}}
\affil{$^1$ Department of Astronomy \& Steward Observatory, University of Arizona, 933 N. Cherry Ave., Tucson, AZ 85721, USA}
\affil{$^2$ Adjunct Astronomer at the National Radio Astronomy Observatory, USA}
\affil{$^3$ Jet Propulsion Laboratory, USA}
\affil{$^4$ Jeremiah Horrocks Institute, University of Central Lancashire, Preston PR1 2HE, UK}
\affil{$^5$ Leiden Observatory, Leiden University, PO Box 9513, 2300 RA, Leiden, The Netherlands}
\affil{$^6$ Kyung Hee University, South Korea}
\affil{$^{7}$ The Dunlap Institute for Astronomy and Astrophysics, University of Toronto, 50 St. George St., Toronto, ON M5S 3H4, Canada}
\affil{$^{8}$ National Science Foundation, USA}
\affil{$^{9}$ National Radio Astronomy Observatory, 520 Edgemont Rd., Charlottesville, VA 22903, USA}

\begin{abstract}

We present deep NH$_3$ observations of the L1495-B218 filaments in the Taurus molecular cloud covering over a 3 degree angular range using the K-band focal plane array on the 100m Green Bank Telescope. The L1495-B218 filaments form an interconnected, nearby, large complex extending over 8 pc. We observed NH$_3$ (1,1) and (2,2) with a spectral resolution of 0.038 km/s and a spatial resolution of 31$''$. Most of the ammonia peaks coincide with intensity peaks in dust continuum maps at 350 $\mu$m and 500 $\mu$m. We deduced physical properties by fitting a model to the observed spectra. We find gas kinetic temperatures of 8 $-$ 15 K, velocity dispersions of 0.05 $-$ 0.25 km/s, and NH$_3$ column densities of 5$\times$10$^{12}$ $-$ 1$\times$10$^{14}$ cm$^{-2}$. The CSAR algorithm, which is a hybrid of seeded-watershed and binary dendrogram algorithms, identifies a total of 55 NH$_3$ structures including 39 leaves and 16 branches. The masses of the NH$_3$ sources range from 0.05 M$_\odot$ to 9.5 M$_\odot$. The masses of NH$_3$ leaves are mostly smaller than their corresponding virial mass estimated from their internal and gravitational energies, which suggests these leaves are gravitationally unbound structures. 9 out of 39 NH$_3$ leaves are gravitationally bound and 7 out of 9 gravitationally bound NH$_3$ leaves are associated with star formation. We also found that 12 out of 30 gravitationally unbound leaves are pressure-confined. Our data suggest that a dense core may form as a pressure-confined structure, evolve to a gravitationally bound core, and undergo collapse to form a protostar.
\end{abstract}

\section{INTRODUCTION}

Filaments are common structures in molecular clouds (e.g. Ridge et al. 2006; Andr{\'e} et al. 2010, 2013; Men'shchikov et al. 2010). They are important structures in star formation because most dense cores form within filaments and stars form within dense cores. Studies of isolated dense cores suggest that stars form mainly as a result of gravity within dense cores, but the formation of filaments and the evolution from filaments to dense cores are still unclear. Recent theoretical studies of molecular clouds suggest that a filament may form by colliding flows of turbulent media and some dense parts further undergo contraction by either self-gravity or ram pressure of colliding flows to make dense cores (e.g. Balsara et al. 2001; Padoan et al. 2001; Klessen et al. 2005; V\'{a}zquez-Semadeni et al. 2005; G\'{o}mez et al. 2007; Nakamura \& Li 2008; Gong \& Ostriker 2009, 2011; Heitsch et al. 2008, 2009; Chen et al. 2014). Since filaments are formed through colliding flows in this scenario, young and less massive dense cores within the filaments are gravitationally unbound, while evolved and massive dense cores tend to be gravitationally bound. Thus, observationally studying the physical state of dense cores within filaments with respect to their evolutionary stage is an essential step to understand how low-mass star formation proceeds within filaments.

The Taurus molecular cloud is one of the best testbeds for studies of both filaments and dense cores because it is nearby (D $\sim$ 140 pc, Loinard et al. 2007, Torres et al. 2009) and forming low-mass stars (no O and B stars) in a relatively isolated mode (with little spatially confusion, see review by Kenyon, G\'{o}mez, \& Whitney 2008). The L1495-B218 filaments are prominent in the Taurus molecular cloud extending over 3 degrees on the sky. They are known to contain cores that range from chemically young starless cores (e.g. L1521B and L1521E; Hirota et al. 2004, Tafall \& Santiago 2004) to more evolved protostellar cores (in B7/L1495; see Rebull et al. 2010). In addition, this region is well mapped in dust extinction (Schmalzl et al. 2010), dust continuum emission (\textit{Herchel Space Observatory}, Palmeirim et al. 2013), dust polarization (Chapman et al. 2011), low density molecular gas tracers ($^{12}$CO and $^{13}$CO Goldsmith et al. 2008; C$^{18}$O Onishi et al. 1996; Hacar \& Tafalla 2013.), and a dense gas tracer (N$_2$H$^+$, Hacar et al. 2013). These maps provide basic information on the structure of the filaments and the positions of some of the densest cores.

Hacar \& Tafalla (2013) observed the L1495-B218 filaments in C$^{18}$O $J$ = 1$-$0 and N$_2$H$^+$ $J$ = 1$-$0. C$^{18}$O $J$ = 1$-$0 is a moderate density ($n_{\rm eff}\leq$ 10$^{3}$ cm$^{-3}$ at 10 K)\footnote{$n_{\rm eff}$ is the effective excitation density, which is needed to produce T$_R$ = 1 K line (Evans 1999; Shirley 2015). See Reiter et al. (2011) for a summary of effective excitation densities.} gas tracer and is a good tracer with which to study velocity structure of moderate density regions since its optical depth is much smaller than those of $^{12}$CO and $^{13}$CO. Using C$^{18}$O, they estimated the LSR velocities and found that the L1495-B218 filaments are not a single coherent velocity structure but a complex group of at least 35 interwoven filaments. However, C$^{18}$O does not trace dense cores because CO is depleted onto dust grains at a high density and low temperature (e.g. Kuiper et al. 1996; Willacy et al. 1998; Kraemer et al. 1999; Caselli et al. 1999; Bacmann et al. 2002; Bergin et al. 2002; Tafalla et al. 2002; Pontoppidan 2006 ; Ford \& Shirley 2011). To find dense cores, the authors used N$_2$H$^+$ $J$ = 1$-$0, which is a dense gas tracer with an effective excitation density above 10$^4$ cm$^{-3}$ (Reiter et al. 2011), but the densities that N$_2$H$^+$ traces are much higher than those traced by C$^{18}$O, which suggests that N$_2$H$^+$ does not trace the complete population of both young and evolved dense cores within the filaments. In order to identify the complete population of dense cores and probe the transition between filaments and dense cores, observations that trace gas with densities intermediate between those traced by C$^{18}$O and N$_2$H$^+$ are required.

Ammonia is an excellent probe that has an effective excitation density intermediate between C$^{18}$O and N$_2$H$^+$. Ammonia starts to form at an early stage of prestellar core evolution and becomes brightest during the later stage of prestellar core evolution (Suzuki et al. 1992). NH$_3$ traces densities $>$10$^3$ cm$^{-3}$ (Flower et al. 2006; Di Francesco et al. 2007). Owing to the NH$_3$ chemistry, the ammonia abundance steeply increases as a molecular cloud core becomes denser and it may be the most abundant molecule after CO that is easily observed within dense cores (Tafalla et al. 2004; Aikawa et al. 2005). Ammonia is also a good tracer for probing the physical condition of filaments. The inversion transitions of ammonia are known to provide an accurate measure of gas kinetic temperature along the line-of-sight (e.g. Ho \& Townes 1983; Rosolowsky et al. 2008) since a transition between meta-stable energy levels is possible only through collisions. If the signal to noise ratio is large enough, the uncertainty of a kinetic temperature measured using the inversion transitions of ammonia may be as low as 0.1 K. An accurate measurement of the kinetic temperature also permits an accurate determination of non-thermal motions, making ammonia an excellent tracer for investigating kinematics.

We observed the L1495-B218 filaments in NH$_3$ (1,1) and $(2,2)$. The complete maps spans over three degrees on the sky and is the largest NH$_3$ map published to date. In this paper, we present observations of filaments and dense cores traced by the ammonia emission. We analyze the physical properties of nested structure identified in the NH$_3$ emission. The kinematics will be explored in a subsequent paper. The layout of this paper is as follows: \S2 briefly introduces our mapping of the L1495-B218 filaments in NH$_3$ and the data reduction procedures. In \S3 we present the basic physical properties of the L1495-B218 filaments and the dense structures identified by the CSAR algorithm within the filaments. In \S4 we discuss physical states of the dense structures identified by the CSAR algorithm. Finally, in \S5 we summarize our results.

\section{OBSERVATION \& DATA REDUCTION }
\subsection{Mapping of the L1495-B218 Filaments in NH$_3$ (1,1) and (2,2)}

We observed the L1495-B218 filaments in the Taurus cloud using the 100 meter Robert C. Byrd Green Bank Telescope (GBT12A295)\footnote{The National Radio Astronomy Observatory is a facility of the National Science Foundation operated under cooperative agreement by Associated Universities, Inc.}. Observations were carried out from April 2012 to February 2013 in 13 shifts totalling 66 hours. We used the K-band focal plane array (KFPA), which has 7 beams each with dual polarization in a hexagonal arrangement (for technical properties of the KFPA, see https://science.nrao.edu/facilities\\/gbt/observing/GBTog.pdf). We used the 7+1 mode of the KFPA, which allows the center beam to observe two frequencies simultaneously while the other 6 beams observe only a single frequency. The bandwidth of each frequency window is 50 MHz. For 7 beams, we set the rest frequency to be 23.702 GHz, which between the strongest hyperfine component of NH$_3$(1,1) (23.6944955 GHz; Lovas \& Dragoset 2003) and NH$_3$(2,2) (23.7226333 GHz; Lovas \& Dragoset 2003). For the central beam we selected a second frequency to include the CCS N$_J$ = 2$_1$ $\rightarrow$ 1$_0$ transition (22.344033 GHz; Yamamoto et al. 1990); these results will be reported in a separate paper. The spectrometer is configured to have 16384 channels across 50 MHz, yielding a resolution of 3.05 KHz (0.0386 km s$^{-1}$ at 23.705 GHz). Since the typical thermal broadening of starless cores is $< 0.2$ km s$^{-1}$ (e.g. Jijina et al. 1999; Caselli et al. 2002; Rosolowsky et al. 2008), the line profiles of NH$_3$ are well resolved in this setup.

We observed the L1495-B218 filaments in 20 rectangular regions that are arranged to cover high extinction regions (Schmalzl et al. 2010) while minimizing the total observation time. Each rectangular region is 20$'$ by 6$'$ or 6$'$ by 20$'$ oriented in Galactic coordinates. Data for each rectangular region are obtained by on-the-fly mapping of 25 rows 20$'$ long with a row spacing of 15.12$''$. The integration time per row is 150 seconds. The resulting integration time for a direction observed with all 7 beams is $\sim$30 seconds, the total observation time for each rectangular region is 1 hour and 20 minutes. The reference OFF positions are picked to be where the extinction is low enough to suggest that they would have negligible NH$_3$ emission, and be close to each rectangular region. The average of two observations of the OFF position made before and after mapping each rectangular region is subtracted from each on-source observation. We also carried out ``daisy" pattern observations to improve the signal-to-noise ratio (SNR) toward 29 bright NH$_3$ regions that we found from the mapping of rectangular regions. Each daisy pattern observation was made with a radius of 2.8$'$ and 22 cycles following a rose curve. The daisy pattern observation completely covers the inner 2.8$'$ radius with all 7 beams for an average integration time of 38 seconds. The total integration time for each daisy pattern is 20 minutes. We updated the pointing of the telescope using the quasar 0403+2600 between every mapping of a rectangular region or every three daisy patterns. The pointing corrections were usually less than 10$''$, compared to the beam size of the GBT at 23 GHz of 31$''$.

Data reduction was performed using the GBT KFPA pipeline. The data are regridded into regular spaced pixels using $AIPS$ and the gain of the 7 beams are calibrated through observing Venus and Jupiter in every observing shift. As a double check on the absolute flux calibration, we also observed the peak dust continuum position of the L1489PPC starless core ($\alpha$ = 04:04:47.6, $\delta$ = +26:19:17.9, $J$2000.0) every observing shift (Young et al. 2004; Ford \& Shirley 2011). The calibrations using Venus and Jupiter and the calibration using L1489PPC agreed within 25\%.

\section{RESULTS}

\subsection{L1495-B218 Filaments Seen in Ammonia}
\subsubsection{Integrated intensity of NH$_3$ (1,1) \& (2,2) and Definition of Subregions}

\begin{figure*}
\includegraphics[angle=90,scale=1]{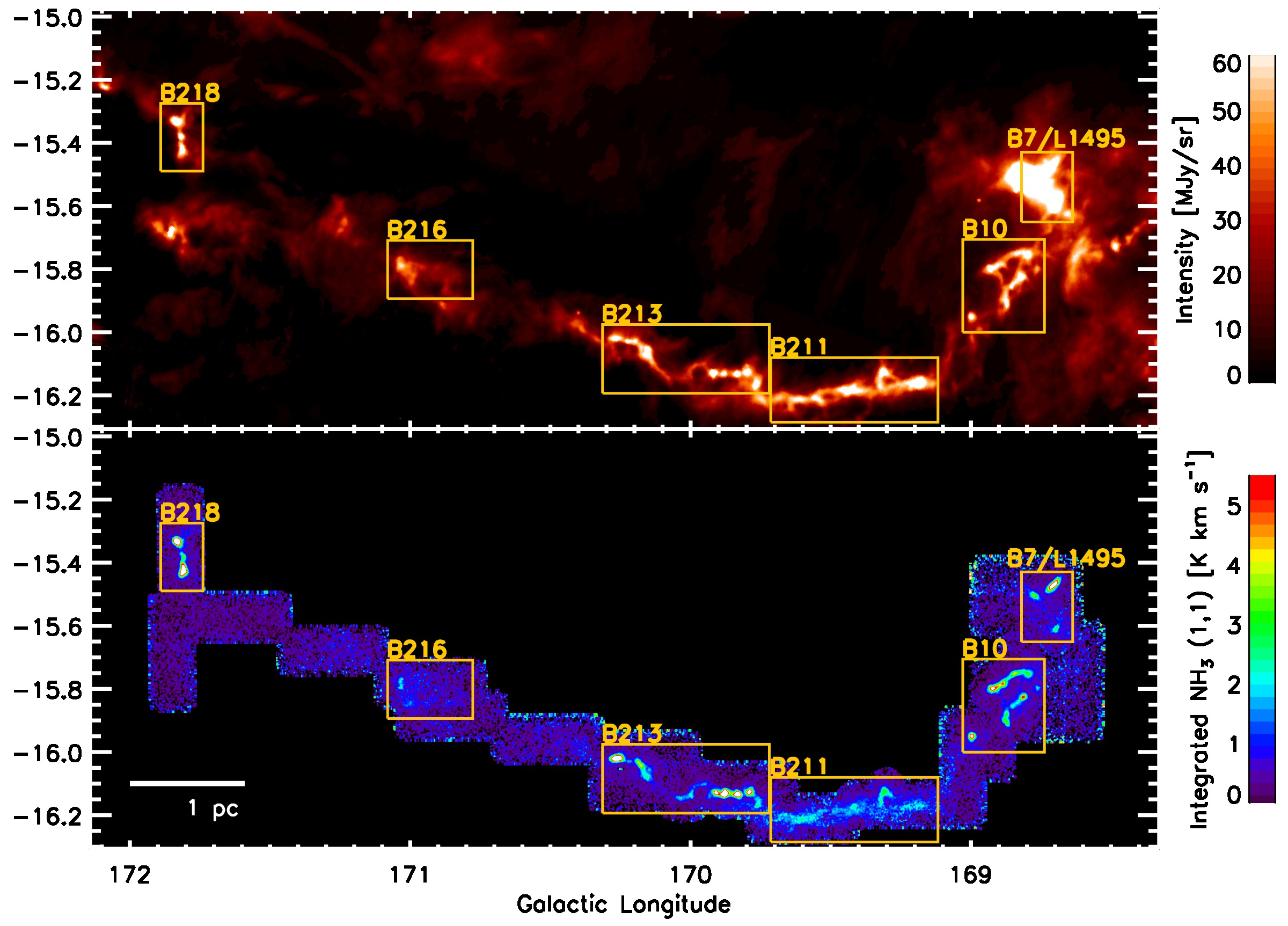}
\caption{Top: 500 $\mu$m dust continuum emission (color) seen by the $SPIRE$ instrument of the {\it Herschel Space Observatory}. Bottom : map of integrated intensity of NH$_3$ (1,1).}
\label{fig1}
\end{figure*}

Figure \ref{fig1} presents the integrated intensity of NH$_3$ (1,1) emission. The NH$_3$ (1,1) map clearly reveals dense cores associated with the L1495-B218 filaments. On the other hand, the filamentary structures between dense cores are barely detected, while those regions are very clearly visible in an extinction map (Schmalzl et al. 2010) and $^{13}$CO (Goldsmith et al. 2008; Pineda et al. 2010b). This suggests that the density of filamentary structures between dense cores may not be high enough to form NH$_3$ efficiently (n$_{\rm H_2}$ $<$ 10$^3$ cm$^{-3}$; Flower et al. 2006) or that ammonia molecules exist but excitation conditions are insufficient to observe the (1,1) emission at our average $rms$ level of 120 mK (Figure 2). The NH$_3$ (2,2) emission is detected only from dense cores and is much weaker than (1,1) emission, which implies that the kinetic temperature may be too low to excite to the $J$=2, $K$=2 state except in dense gas of n$_{\rm H_2}$ $>$ 10$^4$ cm$^{-3}$ (Ho \& Townes 1983; Danby et al. 1988; Maret et al. 2009).

We compare our NH$_3$ (1,1) emissions with 500 $\mu$m dust continuum emission seen by the $SPIRE$ instrument of the {\it Herschel Space Observatory} in Figure \ref{fig1} (Palmerim et al. 2013; Marsh et al. 2014). The 500 $\mu$m dust continuum emission is thermal emission from dust grains and traces the column density structure of the filaments. The bright regions of 500 $\mu$m and NH$_3$ (1,1) emission show a good agreement with each other. The 500 $\mu$m emission is more spatially extended than the NH$_3$ (1,1) emission. The lowest 500 $\mu m$ intensity levels at which NH$_3$ (1,1) and (2,2) are detected at is 25 MJy/sr (N$_{\rm H_2}$ = 5.6 $\times$ 10$^{21}$ cm$^{-2}$ at 10 K) and 177 MJy/sr (N$_{\rm H_2}$ = 3.9 $\times$ 10$^{22}$ cm$^{-2}$ at 10K), respectively. This shows that NH$_3$ is a good tracer with which to identify intermediate to dense regions within filaments.

In Figure \ref{fig5}, we zoom into each region and show the location of protostars identified using $IRAS$ and the {\it Spitzer Space Telescope} (Rebull et al. 2010, 2011; Kryukova et al. 2012). We divide the map into subregions based on Barnard's identification (Barnard 1927). We discuss each region and determine if it is more evolved (having a Class I/flat spectra protostar associated with NH$_3$ emission) or less evolved below.

B7/L1495 contains multiple Class II and III protostars, two flat spectra protostars, and one Class I protostar, none of which is embedded within NH$_3$ core but there are nearby NH$_3$ cores. The three dense cores seen in NH$_3$ are mostly commonly referred to as L1495A-S, L1495A-N, and L1495B (Lee et al. 2001). In B213 (also referred to as L1521D, Codella et al. 1997), there are two Class I protostars embedded in NH$_3$ cores in the eastern part of the region and another three Class I protostars, in the western part of the region. Therefore, B7 and B213 are the most evolved regions in the L1495-B218 filaments and B7 is likely to be more evolved than B213 (see Hacar \& Tafalla et al. 2013).

B218 has only one Flat spectra protostar embedded within a NH$_3$ core. Thus, we classify B218 as an evolved region with star formation, although the NH$_3$ emission is dominated by two bright starless cores. This region is also sometimes confusingly referred to as B217 (Benson \& Myers 1989). We use the notation B218 in this paper since it is closer to the original Barnard catalog position of B218 (Kenyon et al. 2008).

B10 also has only one protostar, which is Class II. In this work, we assume that B10 is a less evolved region because there is no other protostar except the one Class II protostar which is not directly associated with any dense cores in the region.

In contrast to the other regions, B211 and B216 show weak NH$_3$ (1,1) emission and do not possess any known protostars. Thus, both regions are less evolved. B216 region is also named L1521B (Lee et al. 2001) and contains starless cores with large abundances of carbon chain molecules (Hirota et al. 2004).

Hacar \& Tafalla (2013) have also determined the relative ages of subregions using N$_2$H$^+$ 1$-$0 emission and the locations of protostars. Their estimates of relative age of subregions agrees well with ours except for B216. In their observation of N$_2$H$^+$ 1$-$0, they did not detect emission in B216. B216 has weaker NH$_3$ (1,1) emission and stronger CCS emission than B211 (Seo et al. in prep.) suggesting that B216 is a younger region. Thus, B211 and B216 are both relatively young and inactive parts of the L1495-B218 filaments with B216 likely being the most newly condensed part of the filaments.

\subsubsection{Spectral line fitting}

To understand the physical properties of NH$_3$ structures in the L1495-B218 filaments, we fit a simple NH$_3$ spectra model to the observed spectra and deduce physical quantities including kinetic temperature T$_{\rm k}$, excitation temperature T$_{\rm ex}$, total optical depth of NH$_3$ (1,1) emission $\tau_1$, dispersion of emission line $\sigma_v$, and LSR velocity $v_{lsr}$ (See Appendix A for details of the fitting algorithm). The fitting algorithm includes only one velocity component, and this is satisfactory for all regions in the L1495-B218 filaments except B211 (see section 3.2.2 for the LSR velocity of NH$_3$). We only fitted spectra where the detection of NH$_3$ (1,1) and (2,2) emission is above 5$\sigma_{\rm T_{mb}}$ and 3$\sigma_{\rm T_{mb}}$ respectively, where $\sigma_{\rm T_{mb}}$ is the baseline $rms$.

\subsubsection{Gas kinetic temperature and excitation temperature}

Maps of the gas kinetic temperature deduced from NH$_3$ (1,1) and (2,2) emission are presented in Figure \ref{fig7}. The gas kinetic temperature ranges from 8 K to 15 K in most of the region mapped. At the bright NH$_3$ peaks indicated by red arrows, the kinetic temperature decreases toward the core center. The drop in the kinetic temperature is about -1 to -3 K, or about 10\% $-$ 30\% of the average kinetic temperature of surroundings. This temperature drop may ultimately break the dynamical stability of these dense cores and lead them to collapse (see Khesali et al. 2013).

Figure \ref{histograms} presents histograms of physical quantities to investigate statistical differences between subregions. In the first panel of Figure \ref{histograms}, we show histograms of the gas kinetic temperature. Medians of the gas kinetic temperature in subregions are all around 9.5 K with less than 1 K deviation (60\% of all spectra have kinetic temperature within $\pm$1 K of 9.5 K.). The two less evolved regions, B10 and B216, have lower kinetic temperature than the other more evolved regions (we exclude B211 from discussion because it has multiple velocity components and our NH$_3$ fitting model works only for a single velocity component.). The histograms of the gas kinetic temperature demonstrate a tendency that a less evolved region is cooler than a more evolved region and that a denser region is cooler than diffuse region if there are no embedded protostars; however, the gas kinetic temperature differences are subtle.

Histograms of NH$_3$ excitation temperature are shown in the second panel of Figure \ref{histograms}. Median values of the excitation temperature are mostly around 5 K, which is only half of the median values of the kinetic temperature. This demonstrates that NH$_3$ in all regions is subthermally excited.


\subsubsection{LSR velocity}

NH$_3$ lines at 39 NH$_3$ integrated intensity peaks in the L1495-B218 filaments are shown in Figures \ref{vlsr1} and \ref{vlsr2} (for peak locations, see \S3.2.1). The range of NH$_3$ $v_{lsr}$ is from 5.0 km s$^{-1}$ to 7.2 km s$^{-1}$, which is the same range seen in the C$^{18}$O $J$ = 1$-$0 spectra of Hacar et al. (2013). NH$_3$ structures are confined to a single velocity component except in B211. Comparing NH$_3$ to C$^{18}$O (Hacar et al. 2013) and $^{13}$CO (Goldsmith et al. 2008), we found that all NH$_3$ peaks have corresponding velocity components in $^{13}$CO, while only 28 out of 39 NH$_3$ peaks directly associated with the same velocity components in C$^{18}$O. Some NH$_3$ peaks where NH$_3$ integrated intensity is relatively high ($\geq$40 K km s$^{-1}$, $i.e.$ four core-like structures neighboring in a row in B213) do not have corresponding velocity components in C$^{18}$O and $^{13}$CO emission at those places is also weak. We also found a few NH$_3$ peaks with both NH$_3$ and C$^{18}$O emission but discrepancies in their LSR velocities, which may be a sign of chemical differentiation.

The LSR velocity structure of the L1495-B218 filaments was extensively studied by Hacar et al. (2013) using C$^{18}$O $J$= 1$-$0. They identified 35 filaments using a ``friend of friends" algorithm (FIVe; Hacar et al. 2013) and for each filament they deduced physical properties including mass, dispersion of $v_{lsr}$, nonthermal motion, mean kinetic temperature. (See Table 3 in Hacar et al. 2013). Using dense cores identified from their N$_2$H$^+$ $J$ = 1$-$0 observations, they suggested only 7 out of 35 filaments are ``fertile" meaning they contain a N$_2$H$^+$-bright dense core. However, N$_2$H$^+$ traces gas with density higher than 10$^4$ cm$^{-3}$, which is the central density of relatively evolved dense cores (Lada et al. 2008). It is likely that dense cores at early stages are not fully revealed by N$_2$H$^+$. Ammonia molecules have an intermediate effective excitation density that is between C$^{18}$O $J$ = 1$-$0 and N$_2$H$^+$ $J$ = 1$-$0; therefore, our ammonia map may probe the velocity structure of both developing dense cores as well as later stages.

Matching the LSR velocity of NH$_3$ (1,1) and (2,2) structures to 35 filaments of Hacar et al. (2013), we found NH$_3$ (1,1) emission in filaments 6, 8, 10, 11, 15, 20, 28, 32 (Table 3 of Hacar et al. 2013). Hacar et al. (2013) suggested that filaments 6, 10, 11, 15, 20, 32 and 34 are fertile ({\it i.e.} having dense cores in their N$_2$H$^+$ observation). Filament 34 is not within the field of view of our observation, so it is excluded in this study. Our NH$_3$ observation reveals that two more filaments, 8 and 28, also have NH$_3$ emission. Hacar et al. (2013) argued that a filament tends to be sterile when the line density of filament is less than 15 M$_\odot$ pc$^{-1}$, which is the critical density for an isothermal cylinder at 10 K (Stod\'{o}lkiewicz 1963; Ostriker 1964). Filament 28 has a line density of 24.1 M$_\odot$ pc$^{-1}$ (Hacar et al. 2013) but there was no detection of dense cores in N$_2$H$^+$ $J$ = 1$-$0. With our NH$_3$ observations, we found that there are at least two dense cores in filament 28. On the other hand, filament 8 has a line density of 9.3 M$_\odot$ pc$^{-1}$ (Hacar et al. 2013), which is lower than the critical line density, but we see that there is a relatively bright NH$_3$ core. Since Hacar et al. (2013) estimated the masses of filaments from C$^{18}$O $J$ = 1$-$0, it may be possible that they underestimated the mass due to CO depletion. The filament mass estimated from dust continuum may be larger than mass estimated from C$^{18}$O $J$ = 1$-$0, but the filaments cannot be distinguished along the line of sight, which will result in an overestimation of filament masses. From the fact that filament 8 is relatively bright in NH$_3$ (1,1), it is likely that the line density of 9.3 M$_\odot$ pc$^{-1}$ is the minimum line density of filament 8 and the true line density may be higher than the critical value.

\subsubsection{Velocity dispersion}

Velocity dispersions range from 0.05 to 0.25 km s$^{-1}$ except in B211 (B211 has multiple velocity components which are not pursued in this study). Overall, we found that velocity dispersions, which include both thermal and nonthermal components, tend to be wider in the less evolved regions than those in the more evolved regions. Histograms of the velocity dispersions (Figure \ref{histograms}) show this trend clearly. The median values of velocity dispersions for the least evolved regions, B216, is 0.118$^{-1}$, while the most evolved regions, B7, has a median dispersion equal to 0.104 km s$^{-1}$ (the thermal velocity dispersion of NH$_3$ at 10 K is 0.069 km s$^{-1}$).

Using accurate gas kinetic temperatures deduced from NH$_3$ (1,1) and (2,2), we estimate the velocity dispersions of nonthermal kinetic motions $\sigma_{\rm nt}$ (Figure \ref{sigmav}) as follows:
\begin{eqnarray}
\sigma_{\rm nt}=\sqrt{\sigma_v^2-{kT_k\over \mu_{\rm NH_3} m_h}},
\label{sig_nt}
\end{eqnarray}
where $k$ is Boltzmann constant, $\mu_{\rm NH_3}$ is the molecular weight of ammonia (17 a.m.u.), and $m_h$ is the mass of a hydrogen atom. We find most regions of the filaments are supported mainly by thermal pressure (the fourth panel in Figure \ref{histograms}) while some edges of filaments are dominated by nonthermal support ($\sigma_{\rm nt}$ $\geq$ $\sigma_{\rm thermal}$ = 0.188 km s$^{-1}$ at $T_{\rm k}$ = 10 K and $\mu$ = 2.33). The median ratio of thermal support to nonthermal support ($R_p \equiv c_s^2/\sigma_{\rm nt}^2$) is around five. In addition, the median ratios of thermal support to nonthermal support tends to be larger for more evolved regions (median $R_p$ = 6.3 in more evolved region B7, while $R_p$ = 3.7 in less evolved region B216). This suggests that nonthermal motions are considerably dissipated in more evolved regions and making it easier to initiate star formation, which has been seen toward dense cores in other molecular clouds (e.g. Goodman et al. 1998; Pineda et al. 2010a; Tatematsu et al. 2014).

\subsection{Dense NH$_3$ Structures Within the L1495-B218 Filaments}

\subsubsection{Identifying dense structures in the L1495-B218 filaments}

The L1495-B218 filaments have many dense cores in different evolutionary stages of star formation, ranging from starless to protostellar cores. We use our ammonia observations to identify dense cores in the L1495-B218 filaments using the Cardiff Source-finding AlgoRithm (CSAR, Kirk et al. 2013), which uses both a seeded watershed and dendrogram algorithm to find clumpy structures and their spatial relationships. We used the integrated intensity of the central hyperfine group (the strongest hyperfine group of nitrogen splitting) of NH$_3$ (1,1) to identify dense structures because the central hyperfine group gives the highest signal-to-noise ratio compared to other four hyperfine groups. Parameters in the CSAR algorithm are set to find structures above a 7.5-$\sigma$ integrated intensity level in the (1,1) transition ($\sim$3$\sigma$ in (2,2) transition) and to identify a clump with a size larger than $\theta_{\rm mb}/2\sim15''$ ({\it fullbeam} option) and a 2-$\sigma$ enhancement in intensity from its background. We also used the {\it removeedge} option in order not to have overextended edges for a source. In this study, we use 2-$\sigma$ as an intensity interval to clip intensity peaks instead of using typical value of 3-$\sigma$ because with 3-$\sigma$ intensity interval CSAR algorithm tends to combine two peaks as one elongated clumpy structure. These peaks typically lie on top of nested sources which are many $\sigma$ brighter than their background. The CSAR algorithm also had trouble in excluding intensity spikes at edges of maps which have a high noise level ($\sigma_{T_k}$ $\geq$ 0.3 K). In order to exclude the edges of NH$_3$ maps, we made 12 patches and applied the CSAR algorithm to each patch.

The main reason we use NH$_3$ (1,1) instead of dust continuum is because NH$_3$ emission acts as a volume density filter. From the excitation curve of NH$_3$ (1,1), the effective excitation density varies slowly over the range of gas kinetic temperature from 7 K to 15 K. The excitation temperature of dense structures identified by the CSAR algorithm is typically above 3.2 K, and the effective excitation density for the range of observed gas kinetic temperature and NH$_3$ abundance varies only from 7$\times$10$^2$ cm$^{-3}$ to 1 $\times$10$^3$ cm$^{-3}$. This suggests that NH$_3$ (1,1) emission naturally filters out regions with a density lower than $\sim$ 7 $\times$ 10$^2$ cm$^{-3}$. This may make NH$_3$ structures appear smaller than corresponding dust continuum structures, but the NH$_3$ sources identified by the CSAR algorithm are true structures in volume density whereas analysis in dust continuum sees only an enhancement in total column density.


From the integrated intensity of the central hyperfine group of NH$_3$ (1,1), we found 39 leaves (a leaf: a structure containing an intensity peak) and 16 branches (a branch: a nested group of leaves). Figure \ref{csar} shows the locations of NH$_3$ leaves, and Figure \ref{dendro} shows hierarchy among NH$_3$ leaves and branches in a dendrogram. The dendrogram has no lowest common branch because NH$_3$ (1,1) structures are highly fragmented due to the volume density and chemical filtering. In this study we refer to ``active" leaves as those which have a Class I or Flat spectrum protostar within their lowest-level branches in the dendrogram (a.k.a. root) or within a distance of three times their radius. Note that this definition means that active leaves could be either starless or protostellar. Only 4 leaves have embedded protostars and 17 leaves are classified as active. Hacar et al. (2013) found 19 dense cores in N$_2$H$^+$ 1$-$0, while CSAR finds 39 leaves in NH$_3$ (1,1). This difference may be because NH$_3$ is an intermediate density gas tracer that reveals more cores than the denser gas tracer N$_2$H$^+$ and our NH$_3$ observation ($\theta_{\rm mb}$ $\sim$ 31$''$) have better angular resolution than the N$_2$H$^+$ observation ($\theta_{\rm mb}$ $\sim$ 50$''$).

For the comparison between dust and NH$_3$, we also identified dust cores using CSAR. From the 500 $\mu$m dust continuum data of the {\it Herschel Space Observatory} ($\theta_{\rm mb}$ $\sim$ 36$''$), we found 51 dust leaves. Figure \ref{cores} shows the locations and boundaries of dust leaves. 32 dust leaves coincide with 38 NH$_3$ leaves, while 19 dust leaves do not agree with any NH$_3$ leaf. Nine of the dust leaves contain at least two NH$_3$ leaves. This suggests that there may be multiple dense cores or fragments within a core since NH$_3$ is sensitive to volume density whereas the dust continuum is optically thin and depends on the total column density.

In order to compare how well dust continuum and NH$_3$ agree with each other, we investigate the separation of the dust peak position and the NH$_3$ peak position in a dust leaf associated with a NH$_3$ leaf. Most of NH$_3$ peak positions show agreement with corresponding dust peak positions with a median separation of 18$''$.7, which is slightly larger than half the FWHM beam size. However, there are four dust leaves having a separation between dust and NH$_3$ peaks larger than 30$''$ (the maximum separation is 66$''$.2), and three out of four are associated with a nearby protostar (Class I or Flat spectrum). This suggests a large separation may be due to chemical effect produced by the radiation from a protostar. The fact that most of the NH$_3$ sources agree with the dust sources demonstrates that NH$_3$ sources identified by CSAR may represent dense structures within the L1495-B218 filaments. In the following subsections all of the analysis are done with NH$_3$ sources determined from our NH$_3$ data.

\subsubsection{Size and mass}

The physical properties of NH$_3$ leaves and branches are summarized in Tables \ref{table1}, \ref{table2}, and \ref{table3}. The quoted kinetic temperature and velocity dispersion are the median values estimated from NH$_3$ fittings within NH$_3$ leaves and branches except for NH$_3$ leaf 11. This has a peak NH$_3$ (1,1) line intensity weaker than 5$\sigma$ except at its peak location, so we fit the NH$_3$ line at the peak intensity location.

The column density of molecular hydrogen within each structure is estimated from the 500 $\mu$m continuum of \textit{Herschel Space Observatory} using the following equation.
\begin{eqnarray}
N_{\rm H_2}={S_\nu \over \mu m_h B_\nu(T_{\rm dust}) \kappa_\nu \Omega_{\rm ap}}{m_{\rm gas}\over m_{\rm dust}},
\label{dust_col}
\end{eqnarray}
where $N_{\rm H_2}$ is column density of H$_2$ molecules, $S_\nu$ is observed flux at frequency $\nu$, $\mu$ is the mean molecular weight, $m_h$ is the mass of hydrogen atom, $B_\nu$ is black body radiation at frequency $\nu$, $T_{\rm dust}$ is temperature of a dust grain, $\kappa_\nu$ is opacity of dust grain at frequency $\nu$, $\Omega_{\rm ap}$ is the solid angle used in the observation, and $m_{\rm dust}/ m_{\rm gas}$ is dust-to-gas ratio, which we set to 0.01. In this work, we assume $T_{\rm dust} = T_k$, where $T_k$ is the kinetic temperature deduced from our ammonia observations. $\mu$ is 2.8. We use $\kappa_\nu$ of 5.04 cm$^2$ g$^{-1}$ (OH5), which is the column 5 of ``MNR with thin ice mantle" model from Ossenkopf \& Henning (1994). The peak H$_2$ column density of NH$_3$ sources in the L1495-B218 filaments ranges from 6.0 $\times$ 10$^{21}$ to 4.3 $\times$ 10$^{22}$ cm$^{-2}$, which is similar to the typical dense core column density in other molecular clouds (e.g. Jijina et al. 1999; Rosolowsky et al. 2008).

We define the size of a NH$_3$ source as the radius of a circle that has an area equal to that of a NH$_3$ CSAR leaf contour. In the L1495-B218 filaments, the FWHM from Gaussian fitting may not be a useful representation of the size of a dense core because the FWHM contour of a dense core often overlaps with the FWHM contour of its neighboring dense core. Since CSAR separates neighboring sources through hierarchical structure, we use the area identified by CSAR to estimate the size of a structure. Structures were found on scales ranging from 2$\times$10$^{-2}$ pc (4375 AU) to 0.1 pc (20000 AU).

The masses of NH$_3$ sources are estimated by summing up the column density deduced from 500 $\mu m$ continuum within each NH$_3$ leaf and branch (Table \ref{table2} and black circles in Figure \ref{dendro}). This is a typical method for estimating mass of a dense source since three dimensional morphology is usually unknown. If three dimensional morphology is known or assumed, mass of a local background may be subtracted, which will result a smaller dense source mass than the dense source mass estimated by simply summing up the column density. The resulting masses of NH$_3$ leaves range from 0.05 M$_\odot$ to 4.3 M$_\odot$, which are similar to typical dense core masses observed in other studies (0.1-10 M$_\odot$; Motte et al. 1998, 2001; Schmalzl et al. 2010; Onishi et al. 1996, 2002; Kirk et al. 2013). The masses of NH$_3$ branches range from 0.78 M$_\odot$ to 9.5 M$_\odot$. In addition, we also compared three different methods of determining masses of NH$_3$ leaves in Table \ref{table2} and Figure \ref{cmass} (mass estimated from dust continuum, virial mass shown in \S 4.1, and mass estimated from NH$_3$ column density with the fractional abundance of NH$_3$ relative to hydrogen molecules being 1.5 $\times$ 10$^{-9}$). The masses are strongly correlated with each other  with a Spearmann's rank correlation coefficient of $\rho$ = 0.93. This strong linear correlation suggests that the assumptions (i.e. opacity and temperature) used to derive mass from the dust emission do not have strong systematic variations from the lowest mass to highest mass leaves. It also suggests that the properties derived from NH$_3$ sources identified by CSAR are representative of the properties of dense regions in the L1495-B218 filaments, regardless of whether those regions are traced by dust emission or NH$_3$.

We also compared nonthermal support to thermal support within NH$_3$ leaves identified by CSAR in Figure \ref{cmass}. All of the NH$_3$ sources are dominated by thermal support, with an average value of $R_p$ = $c_s^2/\sigma_{\rm nt}^2$ equal to 4.9, and there is no dependency on $R_p$ $vs.$ mass. This shows that nonthermal motions are efficiently dissipated in dense regions and making it easier to initiate star formation.

\subsubsection{Shape and orientation}

The two dimensional morphology of the NH$_3$ sources is estimated using the moment of integrated intensity of NH$_3$ (1,1) with the center at the location of the intensity peak within a leaf or a branch. From the moment of integrated intensity we derive the principal axes, the axis ratios, and the angles with respect to galactic longitude, which are also summarized in Table \ref{table3}. The distribution of angles of NH$_3$ leaves with respect to their lowest-level branches in dendrogram is shown in the third panel of Figure \ref{chist}. The minimum and the maximum phase angles of NH$_3$ leaves with respect to their lowest-level branches are -49$^\circ$ and 53$^\circ$, respectively. Most of NH$_3$ leaves are well aligned with their lowest-level branches within a range of $\pm$40$^\circ$. Since a lowest-level branch corresponds to a filament, this suggests that NH$_3$ leaves are well aligned with filaments and were likely formed within the filaments.

The axis ratio of the NH$_3$ leaves is estimated as short axis/long axis (Table \ref{table3}). The distribution of axis ratios is shown in the fourth panel of Figure \ref{chist}. The axis ratios range from 0.2 to 1.0 and are mostly around 0.5, which is close to the average value obtained from dense core observations by Myers et al. (1991). In order to deduce the three dimensional morphology of NH$_3$ leaves, we compared the observed axis ratio distribution to the probability distribution of apparent axis ratios of spheroids using a Monte Carlo method (see Appendix B for more details). We found the best-fit axis ratio distribution when the spheroids are prolate and the orientation angles (the angle between the equatorial planes of spheroids and the observer's line of sight) range from -80$^\circ$ to 80$^\circ$ (thick solid line in the fourth panel of Figure \ref{chist}). The parameters of the prolate spheroids of the best-fit model are a mean axis ratio (short axis/ long axis) of 0.40 and a dispersion of 0.11 with $\chi^2$ = 5.92 . We could not fit the distribution model using oblate spheroids within 3-$\sigma$ uncertainty ($\chi^2$ = 9) unless we limit the orientation angles much narrower than the observed distribution of the angles between the axes of NH$_3$ leaves and the axes of filaments (third panel of Figure \ref{chist}). Thus, the NH$_3$ leaves are more likely prolate spheroids than oblate spheroids.

\section{Discussion}

The physical and dynamical properties of dense structures depends on the history of their formation. Theoretical studies of molecular filament and dense core formation in a turbulent environment (the gravoturbulence fragmentation scenario) shows that a molecular cloud fragments and develops local density enhancements by colliding flows of turbulence and the density enhancements are grown by self-gravity to evolve into filaments and dense cores. Dense structures formed following this scenario tends to be pressure-confined in their early evolution stage and proceed to a gravitationally bound state. On the other hand, dense cores formed in a quiescent, magnetized environment are typically gravitationally bound through their evolution since fragmentation typically takes place when a structure loses support against gravity. In addition, three dimensional morphologies and alignments of dense structures with magnetic fields also depend on the history of their formation. In a quiescent environment, magnetic fields plays an important role in shaping and aligning dense structures, while the magnetic fields may not play a major role in determining the properties of dense structures in a turbulent environment. In this section, we discuss the dynamical status, three dimensional morphologies, and statistical properties of dense structures in the L1495-B218 filaments as well as uncertainties in our analysis.

\subsection{Virialization}

We investigate the dynamical states of NH$_3$ leaves using the virial theorem. An equation of the virial theorem without magnetic contributions is
\begin{eqnarray}
{1\over 2}{d^2 I \over dt^2} = W+2(T+\Pi)-\int p\vec{x}\cdot \hat{n} dS
\label{virial}
\end{eqnarray}
where $I$ is the moment of inertia, $W$ is the gravitational energy, $T$ is the kinetic energy of systemic motions, $\Pi$ is the internal energy of thermal and nonthermal motions, and the integral term is the work done by the external pressure $p$. $S$ is the core surface and $\hat{n}$ is the unit vector perpendicular to the core surface. We omitted the terms related to magnetic fields because there is no measurement of magnetic fields within NH$_3$ leaves. There are studies using the virial theorem including all terms except magnetic fields (e.g. Hunter 1979; Seo et al. 2013) but only the gravitational and the internal energies are conventionally used for studying dynamical stability of a dense core in observational studies because it is hard to estimate other terms. We also start with only the gravitational and the internal energies in estimating the virial mass and further check whether or not the pressure term is important in confining dense cores. We assume a time-independent steady state so that $d^2I/dt^2$ $=$ 0.

We assumed a NH$_3$ leaf as a homoeoidal ellipsoid with density falling as $\rho$$\sim$$r^{-\beta}$ (Bertoldi \& McKee 1992). The gravitational energy of a homoeoidal ellipsoid is
\begin{eqnarray}
W=-{3 \over 5}{1-{\beta \over 3} \over 1-{2\beta \over 5}} {GM^2\over R}{ \arcsin e \over e},
\label{pe}
\end{eqnarray}
where $e$ is the eccentricity of homoeoidal ellipsoid, and $R$ is the semi-major axis. The internal energy is estimated by
\begin{eqnarray}
\Pi={3 \over 2}M\bar{\sigma}_v^2,
\label{ie}
\end{eqnarray}
where $\bar{\sigma}_v$ is a representative value of velocity dispersion including thermal and nonthermal motions within a NH$_3$ leaf. It can be either the average, median, or mass-weighted values. In this study, we use median values of $\sigma_v$ within NH$_3$ leaves. Finally, the virial mass when $W=2U$ is given as
\begin{eqnarray}
M_{\rm vir} = 5{1-{2\beta \over 5} \over 1-{\beta \over 3}}{R\bar{\sigma}_v^2 \over G}{ e \over \arcsin e}.
\label{vm}
\end{eqnarray}
We first estimated a virial mass with $\rho\sim r^{-2}$ ($\beta = 2$), which is listed in Table \ref{table2}. The eccentricities of leaves are estimated from the axis ratios of NH$_3$ leaves which are given in Table \ref{table3}. The virial masses of NH$_3$ leaves are shown as red circles in Figure \ref{dendro}. The virial mass is calculated for only the NH$_3$ leaves because the NH$_3$ branches usually have either very elongated or irregular shapes, which makes it difficult to calculate the virial mass with conventional assumptions. We find that 30 leaves out of 39 leaves have observed masses smaller than their corresponding virial masses (the first panel in Figure \ref{chist}), which suggests that most of leaves are gravitationally unbound (the median ratio is about two). 7 out of 9 gravitationally bound leaves are active (red), while only two out of 22 inactive (blue) leaves are gravitationally bound. A virial mass of a homoeoidal ellipsoid with $\rho$$\sim$$r^{-1}$ is 50\% larger than the corresponding virial mass of a homoeoidal ellipsoid with $\rho$$\sim$$r^{-2}$. For $\rho$$\sim$$r^{-1}$, there are only two gravitationally bound leaves, both of which are active leaves.

In order to assess the validity of the virial analysis, we must first carefully analyze the uncertainties. One uncertainty may come from a choice of $\bar{\sigma}_v$. In this study, we chose median values of $\sigma_v$ within NH$_3$ leaves but one may use a mean of $\sigma_v$ or a mass-weighted mean of $\sigma_v$ (mass weighted mean is $\bar{\sigma}_v^2$ = $\int N_{\rm H_2}\mu m_h\sigma_v^2 dA/\int N_{\rm H_2}\mu m_h dA$) within a NH$_3$ leaf. We found that the mean values of $\sigma_v$ are at most 10\% larger (typically 5\%) than the corresponding median values, which may result in at most 20\% increase of corresponding virial mass. The mean values may be larger than the median values because $\sigma_v$ tends to be lower at the center of a NH$_3$ leaf and considerably increase toward edges of the NH$_3$ leaf. Mass weighted means on the other hand are in the range from 95\% to 105\% of the corresponding median $\sigma_v$ values. Thus, uncertainties originating from choices of $\bar{\sigma}_v$ do not strongly affect the conclusion of virial mass analysis since ratios of viral masses to observed masses are quite large with respect to these uncertainties.

The largest uncertainties in estimating the observed mass from 500 $\mu$m continuum data are mainly due to assuming $T_d$ = $T_k$ and uncertainty in dust opacity $\kappa$. Comparing the dust temperature, which is estimated by SED fitting of 70, 160, 250, 350, and 500 $\mu$m dust continuums (Palmeirim et al. 2013), to the gas kinetic temperature deduced from NH$_3$ (1,1) and (2,2), we found that the dust temperature is typically 3 $-$ 5 K higher than the gas kinetic temperature. If the dust temperature is 5 K higher than the gas kinetic temperature, the mass estimated from the dust temperature is 35\% less than the mass estimated with the $T_d$ = $T_k$ assumption and only four active NH$_3$ leaves are gravitationally bound.

Another uncertainty is from the dust-to-gas ratio. In this study, we use 0.01, which is a standard value assumed in a molecular cloud (e.g. Beckwith et al. 1990; Lombardi et al. 2006; Schmalzl et al. 2010; Kirk et al. 2013). However, detailed observations of interstellar dust grains in the diffuse ISM indicates that the dust-to-gas ratio is close to 0.0064 (Draine et al. 2011 and reference therein). This increases observed mass of the leaves about 55\% from the current values. In this case, 20 out of 39 leaves are gravitationally bound and 19 leaves are still gravitationally unbound.

Not knowing three dimensional morphologies of leaves also brings uncertainty in estimating masses of leaves. We estimate the mass of a leaf by simply summing up the column mass of pixels that belongs to the leaf, which corresponds to a cylindrical morphology rather than a spheroidal morphology. This has been a conventional method to estimate the masses of dense structures since we do not know three dimensional morphologies (e.g. Kirk et al. 2013). In order to check the uncertainty from assuming different morphologies in estimating the mass of a dense structure, we estimated the masses of leaves by summing up the column density of pixels within leaves after subtracting the average column density of local backgrounds/branches. We found that the masses of leaves are typically 20 $-$ 50\% less than the original values if we subtract a local background from the leaves in estimating leaf mass and that only two our of 39 leaves are gravitationally bound.

The typical uncertainty in the dust opacity is a factor of few (e.g. Shirley et al. 2011) and this results in quite a large uncertainty in estimating mass since M $\propto$ $1/\kappa$. We use OH5 dust opacity but this is not a direct measurement of dust opacity from molecular clouds but a simulated dust opacity based on a coagulated dust model. Shirley et al. (2011) made a direct measurement of the dust opacity at 450 $\mu$m and 850 $\mu$m in the outer envelope of a low-mass Class 0 core, B335, through a comparison between infrared extinction and submillimeter dust continuum. We compared the dust opacity of Shirley et al. (2011) to OH5 dust opacity at 450 $\mu$m and applied the same uncertainty range of dust opacity to OH5 dust opacity at 500 $\mu$m ($\kappa$ = 5.04$^{3.91}_{-1.10}$ cm$^2$ g$^{-1}$). The uncertainty in the observed mass ranges from 57\% lower to 28\% higher with respect to the observed mass given in Table \ref{table2}. If we take the lower limit of uncertainty in the observed mass, two active NH$_3$ leaves are gravitationally bound, while if we take the upper limit of uncertainty in the observed mass 15 out of 39 are gravitationally bound.

Thus, uncertainties in $\sigma_v$, $T_d$, $m_d/m_g$, geometry, backgrounds, and dust opacity are unlikely to change our conclusion that most leaves are gravitationally unbound. This result suggests that a dense core may firstly form as a gravitationally unbound structure, evolve to a gravitationally bound core, and then undergo collapse to form a protostar. Indeed, 10 out of 15 bound leaves in the extreme opacity limit are in more evolved regions.

\subsection{Pressure-confinement}

External pressure may be a source to confine gravitationally unbound structure. Dib et al. (2007) carried out a detailed virial analysis for dense cores forming in a turbulent environment using a numerical method. They showed that the work done by the external pressure in the virial theorem is comparable to the other terms such as internal energy and gravitational energy (Dib et al. 2007). Since most of the NH$_3$ leaves are not gravitationally bound in our simple virial analysis, we checked whether or not work done by the external pressure is comparable to the internal and gravitational energy. The work done by external pressure is estimated as follows
\begin{eqnarray}
\int p\vec{x}\cdot \hat{n}dS = {4 \over 3\sqrt{1-e^2}}\min\{N_{\rm H_2} \mu m_h \sigma_v^2\}_C A,
\label{pw}
\end{eqnarray}
where $A$ is the projected area of NH$_3$ leaf. $N_{\rm H_2} \mu m_h \sigma_v^2$ is the integration of pressure along the line of sight, where $N_{\rm H_2}$ is column density estimated from 500 $\mu$m and $\sigma_v$ is velocity dispersion of molecular gas with $\mu$ = 2.33. We chose the minimum value at the circumference of NH$_3$ leaf, $\min\{N_{\rm H_2} \mu m_h \sigma_v^2\}_C$, because it corresponds to the minimum external pressure exerted on a NH$_3$ leaf. The coefficient of $4/(3\sqrt{1-e^2})$ is for spheroid geometry when one of the principal axes is aligned with the observer's line of sight. We estimate the work done by external pressure using $\min\{N_{\rm H_2} \mu m_h \sigma_v^2\}_C$ because the external pressure is not well known. The estimated work done by external pressure ranges from 5 $\times$ 10$^{40}$ erg to 1 $\times$ 10$^{42}$ erg. In a majority of leaves, gravitational energies are larger than the corresponding work done by external pressure except for 12 leaves (the second panel of Figure \ref{chist}), which suggests that those 12 leaves are pressure-confined. Active leaves tend to have a lower average ratio than that of inactive leaves, which indicates the gravity becomes more important than external pressure as a dense core evolves. Moreover, in 9 out of 12 leaves, the ratios of the work done by external pressure to the gravitational energy are larger than that of the critical Bonnor-Ebert sphere (black dashed vertical line). Thus, these results suggest that the youngest condensations are pressure confined.

A source of the external pressure may be investigated by comparing the internal pressure of the NH$_3$ leaves and the pressure of the surrounding media. The mean internal pressure of NH$_3$ leaves is $<p/k>$ = 2 $\times$ 10$^6$ K cm$^{-3}$, where $p$ is roughly estimated as
\begin{eqnarray}
p = {M\bar{\sigma}_v^2 \over V},
\label{pressure1}
\end{eqnarray}
where $V$ is the leaf volume and $\bar{\sigma}_v$ is the median value of $\sigma_v$ within a NH$_3$ leaf. We assumed that leaf volume is the same as the volume of a prolate spheroid having the observed axis ratio (see \S 4.2.3 for shapes of NH$_3$ leaves). If we assume the density structure of a leaf resembles the critical Bonnor-Ebert sphere, the surface pressure is about 40\% of the mean internal pressure, $<p_{\rm surface}/k>$ = 8 $\times$ 10$^5$ K cm$^{-3}$. This is significantly higher than the gas pressure of the interstellar medium ($<p_{\rm ISM}/k>$ $\approx$ 10$^4$ K cm$^{-3}$, Bertoldi \& McKee 1992; Draine 2011) and still a bit higher than the interstellar turbulence ram pressure, $<p_{\rm ram}/k>$ $\approx$ 5 $\times$ 10$^4$ K cm$^{-3}$ (Lada et al. 2008).

Another source of external pressure may be the weight of the filament material outside of NH$_3$ leaves since the self-gravity of the filaments produces a gravitational potential well and the weight of filaments material outside of NH$_3$ leaves exerts a confining pressure on the NH$_3$ leaves. We estimated the pressure due to the weight of the filament material outside of the NH$_3$ leaves as (Bertoldi \& McKee 1992; Lada et al. 2008):
\begin{eqnarray}
p_{\rm filaments} = {3\pi \over 20}aG\Sigma^2,
\label{pressure2}
\end{eqnarray}
where $\Sigma$ is the mean mass surface density of material outside of leaves, and $a$ is a correction factor for the non-spherical geometry of the filaments. The values of $a$ range from 1.0 to 3.3 depending on the eccentricities and density structures of the filaments (for $a$ = 3.3, we took the maximum axis ratio of the NH$_3$ branches, max(Z/R) $\sim$ 10, and a density structure of $\rho\sim r^{-2}$). The mean of the pressure of the filaments due to their weight is $<p_{\rm filament}/k>$ = 6 $\times$ 10$^5$ K cm$^{-3}$, which is a bit smaller than the surface pressure estimated from Equation (\ref{pressure1}) but within a factor of two. This suggests that the weight of the filaments is one of main sources that confines gravitationally unbound leaves.

Ram pressure due to the inflow of materials onto the filaments and dense cores from the molecular cloud may be another source of confining dense structures. Theoretical studies demonstrated that ram pressure due to mass accretion to dense structures may provide considerable confining pressure (Heitsch 2013a, 2013b) if a converging flow or an inflow is relatively isotropic (a non-isotropic converging flow may disrupt a dense structure, Did et al. 2007). $^{12}$CO and $^{13}$CO observations of the L1495-B218 filaments show that there is a difference of 2 km s$^{-1}$ in the LSR velocities from the north-east to the south-west of B211 (Goldsmith et al. 2008; Palmeirim et al. 2013). This is similar to an infall velocity predicted in a similarity solution of gravitational infall onto a cylinder (Kawachi \& Hanawa 1998). If we accept that the difference in the LSR velocity is due to relatively isotropic inflow of materials from the molecular cloud to the filaments, the ram pressue due to an inflow motion is
\begin{eqnarray}
p_{\rm accretion} = \rho_{\rm p}(R) \delta V^2,
\label{pressure3}
\end{eqnarray}
where $\rho_{\rm p}$ is the density of the best-fit Plummer model to the filament (Palmeirim et al. 2013), and $R$ is the outer radius of the filament which is 0.4 pc for the B211-B213 regions. $\delta V$ is the difference in the LSR velocity across B211 and B213 regions, which is about 2 km s$^{-1}$. The ram pressure due to the accretion of materials from the molecular cloud to filaments are $<p_{\rm filament}/k>$ $\simeq$ 4.5 $\times$ 10$^5$ K cm$^{-3}$, which is smaller than the surface pressure estimated from Equation (\ref{pressure1}) but still within a factor of two. Since there is projection effect in measuring the true inflow velocity, this may be a lower limit of ram pressure due to an inflow motion. Thus, ram pressure due to inflow of materials from the molecular cloud to the filaments may also be one of main sources of pressure that confine gravitationally unbound leaves.

\subsection{Aligned prolate dense structures}

The formation of a filament or a dense core through the gravoturbulence fragmentation scenario predicts that a dense structure may not be strongly aligned with magnetic fields (Nakamura \& Li 2008) unless the magnetic field pressure is considerably stronger than the ram pressure of turbulence. On the other hand, formation through a quasi-static gravitational contraction may result in a strong alignment of the dense core with magnetic fields because contraction occurs more efficiently along magnetic field lines. Although magnetic field directions are not measured within the NH$_3$ sources, magnetic field directions on filament scales (a few 0.1 pc) have been measured through dust polarization in optical and infrared wavelengths along the L1495-B218 filaments (Chapman et al. 2011 and references therein). In B7 and B10 the magnetic fields are relatively parallel to the long axes of a filament. From B213 to B216 the magnetic fields are perpendicular to the long axes of filaments. Dust polarization in B218 is not measured. Since we do not know the magnetic field directions within dense cores and magnetic field directions are aligned in only some regions, it is difficult to ascertain the role of magnetic field in aligning dense cores.

In the gravoturbulence fragmentation scenario, a dense core forms as an oblate spheroid and quickly proceeds to a prolate spheroid or triaxial shape (e.g. Gong \& Ostriker 2011; Kainulainen et al. 2014). The ratio of oblate dense cores to prolate dense cores depends on the Mach number of turbulence and the evolution time. A system with either a higher Mach number or a younger age shows a higher proportion of oblate dense cores. Since the NH$_3$ leaves are likely to be prolate (see \S3.2.3), the L1495-B218 filaments may have been formed in a low Mach number turbulence ($M$ $\leq$ 2; Gong \& Ostriker 2011) if the filaments are formed through the gravoturbulence fragmentation scenario. The velocity dispersions of filaments in the Taurus estimated from $^{13}$CO observations are indeed low Mach number dispersions that range from 0.15 km s$^{-1}$ to 0.65 km s$^{-1}$ (Panopoulou et al. 2014), corresponding to Mach numbers of 0.65 $-$ 3.2, if we take the kinetic temperature of 15 K (Goldsmith et al. 2008). Alternatively, magnetic fields may also make a dense core to have a non-spherical shape because a dense core tends to contract along the magnetic field lines whereas a direction perpendicular to the field is supported by magnetic pressure. If magnetic pressure is the main force making a dense core's non-spherical shape, the dominant shape is an oblate spheroid (Lizano \& Shu 1989). Since most of the NH$_3$ leaves are likely prolate spheroids, magnetic pressure does not seem to be the main cause of shaping a dense core into an elongated morphology, but gravoturbulence fragmentation may have played important roles in shaping the NH$_3$ leaves.

\subsection{Comparing Size, Mass, and Velocity Dispersion}

Larson proposed that there is an empirical relation between line width and size of a structure in a cloud scale (Larson 1981). To see whether there is such a relation in a sub-filament scale, we compare the mean velocity dispersion of NH$_3$ sources with the NH$_3$ source size in Figure \ref{crel}. The velocity dispersion is almost flat on a scale from 0.005 pc to 0.1 pc (the mean velocity dispersion of molecular gas with $\mu$ = 2.33 is 0.204 km s$^{-1}$), and it is close to the thermal velocity dispersion (0.188 km s$^{-1}$ at 10 K and $\mu$ = 2.33). This demonstrates that nonthermal motions are significantly dissipated on scales less than 0.2 pc in the L1495-B218 filaments resulting in a breakdown in the cloud-scale supersonically turbulent scaling relation.

Observations toward molecular clouds shows that molecular clouds are supersonic turbulent (e.g. Larson 1981; Scalo 1984; Miesch \& Bally 1994; Padoan et al. 1999). In a smaller scale, observations toward molecular cloud cores show that dense cores are typically thermally supported and turbulence is significantly dissipated (e.g. Caselli \& Myers 1994; Goodman et al. 1998; Jijina et al. 1999; Caselli et al. 2002). The dense NH$_3$ structures in this study also have the mean $R_p$ of 4.9 (\S3.2.2). On the other hand, the $R_p$ values measured in C$^{18}$O $J$ = 1$-$0 (Hacar et al. 2013) in the L1495-B218 filaments are close unity, and the mean $R_p$ value measured in $^{13}$CO $J$ = 1$-$0 (Panopoulou et al. 2014) is 0.25. Since C$^{18}$O $J$ = 1$-$0 traces less dense and larger structures than those seen by NH$_3$ (1,1) and $^{13}CO$ $J$ = 1$-$0 traces even more extended regions than those traced by C$^{18}$O $J$ = 1$-$0, the dependency of $R_p$ values on the effective excitation density of the tracer suggests that supersonic turbulence cascades from a cloud scale to a core scale and are considerably dissipated on a scale smaller than molecular filaments ($\leq$0.5 pc) in the Taurus. A more detailed analysis of the velocity dispersion and associated kinematic structure will be presented in Paper II.

Mass-size relations are often pursued to investigate the correlation between density structures of sources and their size. The mass and size distributions of leaves and branches in the L1495-B218 filaments are shown in Figure \ref{crel}. Mass and size show a strong correlation of $M\sim r^{1.9}$ (red solid line for NH$_3$ leaves and branches) (power law indexes of leaves only and that of branches only are close to 2 as well). This is close to a global relation for molecular clouds where the power law index is 2 (Larson 1981), which indicates a nearly constant mean column density. Kirk et al. (2013) found a relation with a power law index of 2.35 in the Taurus from Herschel dust continuum observations. Their analysis includes structures larger than 1 pc, whereas the largest structure in this study is only 0.1 pc. If we only take structures less than 0.1 pc from Kirk et al. (2013), the power law index is shallower than 2.35 and close to 2. Kauffmann et al. (2010) studied mass-size relations resolved in several nearby molecular clouds. They also found that the power law index is close to 2 in the Taurus when size is smaller than 0.1 pc, which agrees with our study.

A power index of 2 suggests that the dense structures in the L1495-B218 filaments are gravitationally unbound and have relatively constant column density while a gravitational bound Bonner-Ebert-like core is expected to have a power law index closer to 1 since $\rho\sim r^{-2}$ for the outer profile of a Bonnor-Ebert-like structure. Our virial analysis also shows that most of NH$_3$ leaves in the L1495-B218 filaments are gravitationally unbound. If the NH$_3$ leaves are formed by colliding flows as shown in the gravoturbulence scenario (e.g. Gong \& Ostriker 2013), density structures are more likely to be shallower than $\rho\sim r^{-2}$.

\section{SUMMARY \& CONCLUSIONS}

In this paper, we present extensive NH$_3$ (1,1) and (2,2) maps of the L1495-B218 filaments extending over 3 degrees on the sky in the Taurus molecular cloud with unprecedented depth (average $rms$ is 120 mK), angular resolution (31$''$), and velocity resolution (0.038 km s$^{-1}$). Using the maps, we study the physical properties of the L1495-B218 filaments by fitting 13109 NH$_3$ (1,1) and (2,2) spectra using an adaptive mesh refinement search of $\chi^2$ space. The main results are following.

1. From our ammonia observations and a protostar catalog (Rebull et al. 2010), we confirm that B211 and B216 are young, less evolved regions, while B213 and B7 are actively star-forming, older, more evolved regions. The young regions have NH$_3$ column density around 1 $\times$ 10$^{13}$ cm$^{-2}$ and do not have any protostars. On the other hand, the more evolved, older regions have NH$_3$ column density up to 1 $\times$ 10$^{14}$ cm$^{-2}$ and have multiple protostars. NH$_3$ emission is mostly confined to a single velocity component and the inversion levels are subthermally populated.

2. Gas kinetic temperatures in the L1495-B218 filaments deduced from NH$_3$ (1,1) and (2,2) lines reveal that the filaments to be very cold (8 $-$ 15 K with median value equal to 9.5 K, 60\% of spectra are in the range of 9.5 $\pm$ 1 K). We found that there is a small difference in gas kinetic temperature between the more evolved regions (B7, B213, and B218) and less evolved regions (B10, B211, and B216); with the more evolved regions having a higher median gas kinetic temperature (by at most 0.5 K) than the less evolved regions. Gas kinetic temperatures tend to decrease toward dense core centers at the level of a few Kelvins. This may ultimately affect the stability and dynamics of dense cores and filaments.

3. The nonthermal velocity dispersion of NH$_3$ (1,1) and (2,2) lines in the L1495-B218 filaments ($\sim$0.08 km s$^{-1}$) is considerably narrower than those of C$^{18}$O $J$ = 1$-$0 ($\sim$0.15 km s$^{-1}$; Hacar et al. 2013) and $^{13}$CO $J$ = 1$-$0 ($\sim$1 km s$^{-1}$; Goldsmith et al. 2008) in the same region. This suggests that nonthermal motions in the regions traced by NH$_3$ have been significantly dissipated compared to those found in the more diffuse portions of the filaments. As a result, dense regions do not display a Larson's size-line width relationship.

4. Results from the CSAR clump-finding algorithm found 39 NH$_3$ peaks (leaves) and 16 nested groups (branches), which is about twice the number that Hacar et al. (2013) found in N$_2$H$^+$. This may be because NH$_3$ reveals lower density cores than does N$_2$H$^+$ and our NH$_3$ observations have 67\% better spatial resolution than the N$_2$H$^+$ data of Hacar et al. (2013). The NH$_3$ leaves and branches are identified on a scale from 0.01 pc to 0.1 pc and have masses ranging from 0.05 M$_\odot$ to 9.5 M$_\odot$.

5. Masses of dense cores derived from dust continuum and NH$_3$ show a strong correlation in the L1495-B218 filaments. This suggests that NH$_3$ is an exceptional tracer for tracing dense cores in these filaments. A similar study across other molecular clouds will be interesting.

6. Most of the NH$_3$ leaves have observed masses smaller than the corresponding virial masses, which means they are likely to be gravitationally unbound structures. We found nine NH$_3$ leaves to be gravitationally bound and that 7 out of 9 leaves are either protostellar or within branches associated with star formation activity, while 30 NH$_3$ leaves are gravitational unbound and only 10 out of 30 unbound leaves are either protostellar or within branches associated with star formation. We also found that 12 out of 30 gravitationally unbound leaves are pressure-confined. This is a similar conclusion that Kainulainen et al. (2011) found in larger scale clumps ($>$0.1 pc) in nearby clouds, while the dense structures in this study are from 0.005 pc to 0.1 pc.

7. Two sources may provide confining pressure to gravitationally unbound NH$_3$ leaves. The self-gravity of the filaments produces a gravitational potential well and the weight of filament materials outside of NH$_3$ leaves may exert a confining pressure on the surface of NH$_3$ leaves. The estimated mean pressure due to weight of filaments is a bit smaller than the surface pressure needed to confine gravitationally unbound leaves but within factor of two. The other source may be ram pressure due to large scale inflows. $^{12}$CO and $^{13}$CO observations show that there may be inflows toward the L1495-B218 filaments. The minimum ram pressure due to the inflows is about half of the surface pressure needed to confine gravitationally unbound leaves.

8. Most of the NH$_3$ leaves are well aligned with their lowest-level branches within the range of $\pm$40 degrees. The distribution of the apparent axis ratios of NH$_3$ leaves is closer to that of randomly oriented prolate spheroids than that of randomly oriented oblate spheroids, which indicates most of the NH$_3$ leaves are likely to be prolate. Since magnetic pressure would shape a dense core to be a oblate spheroid, magnetic fields do not seem to play an important role in determining the observed axis ratio of NH$_3$ leaves.

9. The NH$_3$ leaves and branches each follow a mass radius relationship of M $\sim$ R$^{-1.9\pm0.08}$.

Finally, both core-like and filamentary structures in the L1495-B218 filaments were successfully probed using NH$_3$, which suggests that NH$_3$ is an excellent tracer of $\geq$10$^3$ cm$^{-3}$ gas for probing both dense cores and surrounding filaments in molecular clouds. We found that most of the NH$_3$ structures agree in the LSR velocity with the coherent filaments found by Hacar et al. (2013). We also found that most of dense NH$_3$ structures (the CSAR leaves and branched) coincide with the dynamically critical filaments ($\geq$15 M$_\odot$ pc$^{-1}$) in Hacar et al. (2013). These results suggest that a dense core may form as a pressure-confined structure, evolve to a gravitationally bound state within critical filaments, and then undergo star formation.

\acknowledgments

We are grateful to anonymous referee for helpful suggestions. We are also grateful to A. Hacar and M. Tafalla for providing C$^{18}$O data. Y. Seo was support by a NRAO Observer's Grant (GBT/13A-126) and partially by NSF Grant AST-1008577. Y. Shirley was partially supported by NSF Grants AST-1008577 and AST-1410190. This work was carried out in part at the Jet Propulsion Laboratory, which is operated for NASA by the California Institute of Technology.


\appendix

\section{METHODS OF ANALYZING PHYSICAL PROPERTIES}

We derived physical quantities through finding the minimum reduced $\chi_r^2$ value between the observed NH$_3$ (1,1) and (2,2) spectra and the spectra model. Our method of finding the minimum $\chi_r^2$ value is a bit different from previous work in that we do not use a $\chi_r^2$ minimization method that utilizes a covariance matrix to calculate the uncertainties but instead explore a wide range of parameter space in adaptive grids to obtain the global minimum and determine the shape of $\chi_r^2$ volume as functions of the parameters. The common method of minimizing $\chi_r^2$ value utilizes a covariance matrix because the method is relatively easy and computationally fast, but parameters should be random variables. If a parameter has a specific non-random distribution, the 1-$\sigma$ uncertainty of that parameter may be not symmetric and may differ from the standard deviation estimated by the covariance matrix.

Exploring parameter space gives proper estimation of uncertainties, but it is usually computationally expensive if the parameter space is large and is searched at high resolution. To reduce computational time, we use the adaptive mesh refinement method (AMR method). First, we set up ranges of parameter space to explore $[x_{\rm 0,lower}, x_{\rm 0,upper}]$, where the subscript $0$ denotes zeroth layer, the subscript lower means the lower limit of the parameter space which we are exploring, and the subscript upper means the upper limit. The number of grids in each parameter can be chosen arbitrarily, but for efficiency we used fewer than 20 grids for each parameter. Second, we calculate the minimum $\chi_r^2$ at given grid points and find its location $x_{\rm 0,min}$. Third, we set up new range of parameter space and explore as the first layer $[x_{\rm 1,lower}, x_{\rm 1,upper}]$. The range can be given as
\begin{eqnarray}
x_{\rm 1,lower}=x_{\rm 0,min}-{x_{\rm 0,upper}-x_{\rm 0,lower} \over 2}~{\rm when}~x_{\rm 1,lower} \geq x_{\rm 0,lower}
\end{eqnarray}
and
\begin{eqnarray}
x_{\rm 1,upper}=x_{\rm 0,min}+{x_{\rm 0,upper}-x_{\rm 0,lower} \over 2}~{\rm when}~x_{\rm 1,upper} \leq x_{\rm 0,upper}.
\end{eqnarray}
If $x_{\rm 1,lower}$ is smaller than $x_{\rm 0,lower}$, then $x_{\rm 1,lower}= x_{\rm 0,lower}$. Likewise, if $x_{\rm 1,upper}$ is larger than $x_{\rm 0,upper}$, then $x_{\rm 1,upper}= x_{\rm 0,upper}$. Fourth, we estimate the $\chi_r^2$ value within the new range of parameter space with the same number of grids we used in the first step. At this step, the resolution is double that of the previous layer. Fifth, we repeat the second to the forth steps until we achieve the desired resolution. If there is the global minimum of $\chi_r^2$ value within the range of the zeroth parameter space, this code can find the global minimum with an uncertainty of the resolution of the final layer unless the widths of local minima and global minimum are much narrower than the resolution of the final layer. In our calculations, we set the finest velocity resolution of the model to be at least quarter of the velocity resolution of our observation, so we may not miss the global minimum.

For the model of the NH$_3$ line profiles, we use the model described in Rosolowsky et al. (2008).

Figure \ref{line_examples} shows two examples of fitting the observed NH$_3$ spectra and Figure \ref{chisq_examples} shows the $\chi_r^2$ of the two examples in T$_k$ $vs.$ $\tau_1$ space. The top panel of Figure \ref{line_examples} is the best-fit of the simple emission model to an observed strong NH$_3$ (1,1) emission of $\sim$32$\sigma_{\rm T_{mb}}$ and the left panel of Figure \ref{chisq_examples} is the reduced $\chi^2$ of fitting in T$_k$ $vs.$ $\tau_1$ space with the other three parameters fixed at the best-fit values. The bottom panel of Figure \ref{line_examples} shows an example of a weaker emission spectra of NH$_3$ (1,1) with the peak signal-to-noise ratio (SNR) being $\sim$7.5 and the right panel of Figure \ref{chisq_examples} is the reduced $\chi^2$ of fitting in T$_k$ $vs.$ $\tau_1$ space with the other three parameters fixed at their best-fit values. Uncertainties of physical parameters are about three times larger in the weak line than those in the strong line.

We fit 13,109 spectra which satisfy the detection of NH$_3$ (1,1) and (2,2) emission is above 5$\sigma_{\rm T_{mb}}$ and 3$\sigma_{\rm T_{mb}}$, respectively, where $\sigma_{\rm T_{mb}}$ is the baseline $rms$. The parameter ranges for searching for the best-fit are [7.5 K, 30.0 K] for T$_{\rm k}$, [2.75K, 30K] for T$_{\rm ex}$, [5 $\times$ 10$^{-3}$, 20] for $\tau_1$, and [0.05 km s$^{-1}$, 0.3 km s$^{-1}$] for $\sigma_v$. The $v_{lsr}$ is preliminarily estimated from the peak intensities of NH$_3$ (1,1) and (2,2) lines and a more accurate value is searched for in a range of $\pm$0.1 km s$^{-1}$ from the guessed value. Typical 1-$\sigma$ uncertainties are $\pm$0.77 K for T$_{\rm k}$, $\pm$0.85 K for T$_{\rm ex}$, $\pm$0.26 for $\tau_1$, $\pm$0.013 km s$^{-1}$ for $\sigma_v$, and $\pm$ 0.020 km s$^{-1}$ for $v_{lsr}$. The overall uncertainty distributions of 13,109 fits to the spectra are shown in Figure \ref{errors}.

\section{MODELING APPARENT AXIS RATIO DISTRIBUTIONS OF DENSE STRUCTURES USING A MONTE CARLO METHOD}

We analyzed morphology of dense structures through estimating the minimum $\chi^2$ value (not reduced $\chi^2$) between the observed distribution of apparent axis ratios and distribution models. The distribution models are made using a Monte Carlo method. For each model, we imposed ten million spheroids and assumed all of the spheroids are either in oblate or prolate shapes. The apparent axis ratios of the spheroids are determined by the intrinsic axis ratio of the spheroid and the angle between observer's line of sight and the equatorial plane of the spheroid as follow:
\begin{eqnarray}
p(q, \theta)=[1-(1-q^2)\cos\theta]^{1 \over 2}~{\rm for ~ oblate},  \\
p(q, \theta)=q[1-(1-q^2)\sin\theta]^{1 \over 2}~{\rm for ~ prolate},
\label{pq}
\end{eqnarray}
where $p$ is the apparent axis ratio of the spheroid, $q$ is the intrinsic axis ratio of the spheroid (short axis/long axis), and $\theta$ is the angle between observer's line of sight and the equatorial plane of the spheroid. We assumed that the intrinsic axis ratios of the spheroids follow a Gaussian distribution and the angles between the observer's line of sight and the equatorial planes of spheroids have random values with a uniform distribution within $[-\theta_{\rm limit},\theta_{\rm limit}]$. The mean value and width of a Gaussian distribution and $\theta_{\rm limit}$ serves as input parameters of the model. Using above equations with given input parameters, we modeled a distribution of apparent axis ratios of the spheroids. We made 400,000 models with different mean values $\langle q\rangle$ and widths $\sigma_q$ of Gaussian distributions of the intrinsic axis ratios and $\theta_{\rm limit}$. The range of $\langle q\rangle$ that is exploited in our models is from 0.01 to 1 with an interval of 0.01 and the range of $\sigma_q$ is from 0.01 to 1 with interval of 0.01. The range of $\theta_{\rm limit}$ is from 0 degree to 90 degree with an interval of 5 degree. We compared all 400,000 distribution models to the observed distribution and found the best-fit model by minimizing the $\chi^2$ value.

\clearpage

\clearpage

\newpage
\begin{deluxetable}{ c c c c c c c c c c}
\rotate
\tablecaption{Properties of NH$_3$ sources}
\tablewidth{0pt}
  \startdata
  \hline
ID$^a$	& $b$	& $l$ & $\alpha$ & $\delta$ & Peak NH$_3^{ b}$	   & Peak Dust$^c$ & N$_{H_2}^{ d}$        &  Embedded$^{ f}$ &   Area \\
		&    	&     & $J$2000.0& $J$2000.0& [K km s$^{-1}$]      & [MJy/sr]     & [10$^{22}$ cm$^{-2}$]  &  YSOs    &    [10$^{-2}$ pc$^2$] \\
\hline
1	&	+168:42:22.8  	&	-15:28:12 	&	4:18:33.0	&	+28:27:58	&	 7.07$\pm$0.09	&	85.9	&	1.69$_{-0.14}^{0.11}$	&	N	&	0.046	\\
1.5	&	+168:42:22.8  	&	-15:28:12 	&	4:18:33.0	&	+28:27:58	&	 7.07$\pm$0.09	&	85.9	&	1.69$_{-0.14}^{0.11}$	&	-	&	1.206	\\
2	&	+168:42:59.9  	&	-15:28:42 	&	4:18:33.1	&	+28:27:11	&	 6.67$\pm$0.08	&	91.2	&	2.28$_{-0.17}^{0.15}$	&	N	&	0.012	\\
3	&	+168:46:43.4  	&	-15:30:18 	&	4:18:39.1	&	+28:23:29	&	 3.60$\pm$0.08	&	176.6	&	4.03$_{-0.43}^{0.53}$	&	N	&	0.526	\\
4	&	+168:47:46.1  	&	-15:29:12 	&	4:18:45.9	&	+28:23:30	&	 1.02$\pm$0.13	&	112.4	&	4.30$_{-2.03}^{2.51}$	&	N	&	0.046	\\
5	&	+168:41:41.8  	&	-15:36:18 	&	4:18:04.2	&	+28:22:52	&	 2.82$\pm$0.09	&	94.9	&	2.48$_{-0.29}^{0.73}$	&	N	&	0.473	\\
6	&	+168:47:45.6  	&	-15:45:18 	&	4:17:52.8	&	+28:12:26	&	 3.07$\pm$0.06	&	75.0	&	2.33$_{-0.27}^{0.55}$	&	N	&	0.194	\\
6.5	&	+168:50:46.4  	&	-15:45:06 	&	4:18:02.6	&	+28:10:28	&	 3.83$\pm$0.06	&	58.5	&	1.43$_{-0.16}^{0.14}$	&	-	&	0.507	\\
7	&	+168:50:46.4  	&	-15:45:06 	&	4:18:02.6	&	+28:10:28	&	 3.83$\pm$0.06	&	58.5	&	1.43$_{-0.16}^{0.14}$	&	N	&	0.257	\\
7.5	&	+168:55:25.8  	&	-15:48:00 	&	4:18:07.0	&	+28:05:13	&	 6.41$\pm$0.06	&	70.5	&	2.64$_{-0.27}^{0.17}$	&	-	&	2.881	\\
8	&	+168:53:21.5  	&	-15:47:06 	&	4:18:03.7	&	+28:07:17	&	 5.04$\pm$0.05	&	57.3	&	1.60$_{-0.12}^{0.10}$	&	N	&	0.071	\\
8.5	&	+168:55:25.8  	&	-15:48:00 	&	4:18:07.0	&	+28:05:13	&	 6.41$\pm$0.06	&	70.5	&	2.64$_{-0.27}^{0.17}$	&	-	&	0.768	\\
9	&	+168:55:25.8  	&	-15:48:00 	&	4:18:07.0	&	+28:05:13	&	 6.41$\pm$0.06	&	70.5	&	2.64$_{-0.27}^{0.17}$	&	N	&	0.192	\\
10	&	+168:45:39.7  	&	-15:48:00 	&	4:17:37.6	&	+28:12:02	&	 1.57$\pm$0.10	&	52.5	&	1.57$_{-0.41}^{0.80}$	&	N	&	0.348	\\
11	&	+168:46:00.1  	&	-15:44:12 	&	4:17:51.2	&	+28:14:25	&	 0.82$\pm$0.10	&	39.2	&	1.27$_{-0.52}^{0.92}$	&	N	&	0.050	\\
12	&	+168:48:46.1  	&	-15:49:36 	&	4:17:41.7	&	+28:08:46	&	 5.34$\pm$0.06	&	79.0	&	2.53$_{-0.07}^{0.19}$	&	N	&	0.226	\\
12.5	&	 +168:48:46.1 	&	-15:49:36 	&	4:17:41.7	&	+28:08:46	&	 5.34$\pm$0.06	&	79.0	&	2.53$_{-0.07}^{0.19}$	&	-	&	0.357	\\
13	&	+168:49:48.1  	&	-15:50:24 	&	4:17:42.2	&	+28:07:29	&	 2.08$\pm$0.07	&	58.6	&	1.02$_{-0.21}^{0.12}$	&	N	&	0.017	\\
13.25	&	 +168:48:46.1 	&	-15:49:36 	&	4:17:41.7	&	+28:08:46	&	 5.34$\pm$0.06	&	79.0	&	2.53$_{-0.07}^{0.19}$	&	-	&	1.076	\\
14	&	+168:51:02.6  	&	-15:51:18 	&	4:17:43.0	&	+28:06:00	&	 2.19$\pm$0.08	&	59.4	&	1.51$_{-0.25}^{0.52}$	&	N	&	0.070	\\
15	&	+168:52:22.9  	&	-15:53:06 	&	4:17:41.1	&	+28:03:50	&	 2.87$\pm$0.06	&	57.4	&	2.14$_{-0.31}^{0.28}$	&	N	&	0.050	\\
15.5	&	 +168:52:16.1 	&	-15:54:30 	&	4:17:36.1	&	+28:02:57	&	 3.43$\pm$0.07	&	71.7	&	1.92$_{-0.20}^{0.37}$	&	-	&	0.884	\\
16	&	+168:52:16.1  	&	-15:54:30 	&	4:17:36.1	&	+28:02:57	&	 3.43$\pm$0.07	&	71.7	&	1.92$_{-0.20}^{0.37}$	&	N	&	0.146	\\
17	&	+168:59:50.6  	&	-15:57:06 	&	4:17:50.3	&	+27:55:52	&	 6.17$\pm$0.07	&	70.8	&	2.25$_{-0.18}^{0.23}$	&	N	&	0.655	\\
18	&	+169:18:24.7  	&	-16:07:30 	&	4:18:11.7	&	+27:35:46	&	 2.90$\pm$0.07	&	74.7	&	1.87$_{-0.17}^{0.36}$	&	N	&	0.030	\\
18.5	&	 +169:18:37.1 	&	-16:07:54 	&	4:18:11.0	&	+27:35:21	&	 2.86$\pm$0.07	&	73.4	&	2.21$_{-0.28}^{0.28}$	&	-	&	0.242	\\
19	&	 +169:18:37.1 	&	-16:07:54 	&	4:18:11.0	&	+27:35:21	&	 2.86$\pm$0.07	&	73.4	&	2.21$_{-0.28}^{0.28}$	&	N	&	0.015	\\
19.25	&	 +169:18:37.1 	&	-16:07:54 	&	4:18:11.0	&	+27:35:21	&	 2.86$\pm$0.07	&	73.4	&	2.21$_{-0.28}^{0.28}$	&	-	&	1.161	\\
20	&	+169:19:14.2  	&	-16:09:24 	&	4:18:07.9	&	+27:33:53	&	 2.04$\pm$0.09	&	55.9	&	1.74$_{-0.50}^{0.32}$	&	N	&	0.078	\\
21	&	 +169:45:40.7 	&	-16:10:18 	&	4:19:23.3	&	+27:14:46	&	 2.30$\pm$0.11	&	82.6	&	1.52$_{-0.31}^{0.50}$	&	N	&	0.481	\\
21.925	&	+169:49:44.6  	&	-16:08:06 	&	4:19:42.5	&	+27:13:25	&	 9.49$\pm$0.07	&	120.0	&	2.05$_{-0.07}^{0.08}$	&	-	&	2.979	\\
22	&	 +169:47:27.3 	&	-16:07:42 	&	4:19:37.1	&	+27:15:18	&	 7.40$\pm$0.08	&	69.6	&	1.71$_{-0.10}^{0.15}$	&	N	&	0.610	\\
22.85	&	 +169:49:44.6 	&	-16:08:06 	&	4:19:42.5	&	+27:13:25	&	 9.49$\pm$0.07	&	120.0	&	2.05$_{-0.07}^{0.08}$	&	-	&	2.363	\\
23	&	+169:49:44.6  	&	-16:08:06 	&	4:19:42.5	&	+27:13:25	&	 9.49$\pm$0.07	&	120.0	&	2.05$_{-0.07}^{0.08}$	&	Y	&	0.279	\\
23.75	&	 +169:49:44.6 	&	-16:08:06 	&	4:19:42.5	&	+27:13:25	&	 9.49$\pm$0.07	&	120.0	&	2.05$_{-0.07}^{0.08}$	&	-	&	1.687	\\
24	&	+169:52:39.5  	&	-16:08:00 	&	4:19:51.5	&	+27:11:26	&	10.45$\pm$0.06	&	67.0	&	2.24$_{-0.08}^{0.09}$	&	N	&	0.231	\\
24.5	&	 +169:52:39.5 	&	-16:08:00 	&	4:19:51.5	&	+27:11:26	&	10.45$\pm$0.06	&	67.0	&	2.24$_{-0.08}^{0.09}$	&	-	&	0.524	\\
25	&	 +169:54:32.0 	&	-16:07:48 	&	4:19:57.6	&	+27:10:15	&	 7.36$\pm$0.08	&	74.7	&	1.57$_{-0.08}^{0.07}$	&	Y	&	0.075	\\
26	&	 +169:56:49.5 	&	-16:06:12 	&	4:20:09.7	&	+27:09:44	&	 1.40$\pm$0.08	&	35.4	&	0.83$_{-0.22}^{0.11}$	&	N	&	0.118	\\
27	&	 +169:59:13.1 	&	-16:06:48 	&	4:20:14.7	&	+27:07:39	&	 1.11$\pm$0.09	&	34.1	&	0.69$_{-0.07}^{0.14}$	&	N	&	0.045	\\
27.75	&	 +170:00:03.0 	&	-16:08:18 	&	4:20:12.2	&	+27:06:02	&	 1.88$\pm$0.10	&	33.6	&	0.60$_{-0.11}^{0.18}$	&	-	&	0.605	\\
28	&	 +170:00:03.0 	&	-16:08:18 	&	4:20:12.2	&	+27:06:02	&	 1.88$\pm$0.10	&	33.6	&	0.60$_{-0.11}^{0.18}$	&	N	&	0.133	\\
28.5	&	 +170:00:03.0 	&	-16:08:18 	&	4:20:12.2	&	+27:06:02	&	 1.88$\pm$0.10	&	33.6	&	0.60$_{-0.11}^{0.18}$	&	-	&	0.559	\\
29	&	+170:01:49.1  	&	-16:08:54 	&	4:20:15.5	&	+27:04:23	&	 1.90$\pm$0.09	&	29.6	&	0.74$_{-0.18}^{0.11}$	&	N	&	0.124	\\
30	&	 +170:10:40.2 	&	-16:02:30 	&	4:21:02.5	&	+27:02:30	&	 3.81$\pm$0.09	&	53.9	&	1.29$_{-0.21}^{0.14}$	&	N	&	0.363	\\
30.5	&	 +170:09:25.2 	&	-16:04:42 	&	4:20:51.6	&	+27:01:54	&	 2.23$\pm$0.08	&	59.6	&	1.29$_{-0.17}^{0.51}$	&	-	&	0.927	\\
31	&	+170:09:25.2  	&	-16:04:42 	&	4:20:51.6	&	+27:01:54	&	 2.23$\pm$0.08	&	59.6	&	1.29$_{-0.17}^{0.51}$	&	N	&	0.138	\\
32	&	+170:08:47.8  	&	-16:03:24 	&	4:20:54.1	&	+27:03:13	&	 1.54$\pm$0.07	&	70.2	&	1.87$_{-0.37}^{0.61}$	&	N	&	0.051	\\
33	&	+170:15:58.6  	&	-16:01:24 	&	4:21:21.7	&	+26:59:31	&	 9.36$\pm$0.09	&	73.2	&	2.38$_{-0.10}^{0.21}$	&	N	&	0.838	\\
34	&	+170:13:16.3  	&	-16:01:48 	&	4:21:12.5	&	+27:01:09	&	 1.41$\pm$0.10	&	71.9	&	0.71$_{-0.11}^{0.38}$	&	Y	&	0.041	\\
35	&	+171:02:00.7  	&	-15:47:48 	&	4:24:20.6	&	+26:36:02	&	 2.22$\pm$0.15	&	54.9	&	1.73$_{-0.36}^{1.59}$	&	N	&	0.091	\\
36	&	+171:01:47.9  	&	-15:46:12 	&	4:24:25.3	&	+26:37:16	&	 1.87$\pm$0.12	&	44.3	&	1.43$_{-0.21}^{0.36}$	&	N	&	0.076	\\
37	&	+171:48:05.3  	&	-15:25:24 	&	4:27:47.5	&	+26:17:58	&	 9.33$\pm$0.11	&	60.6	&	1.82$_{-0.11}^{0.15}$	&	N	&	0.639	\\
38	&	+171:48:04.1  	&	-15:22:54 	&	4:27:55.8	&	+26:19:38	&	 3.62$\pm$0.11	&	48.6	&	1.96$_{-0.42}^{0.43}$	&	Y	&	0.090	\\
39	&	+171:49:30.0  	&	-15:20:06 	&	4:28:09.2	&	+26:20:28	&	 9.20$\pm$0.12	&	89.6	&	2.64$_{-0.23}^{0.14}$	&	N	&	0.809	\\
\enddata
\tablenotetext{a}{\footnotesize IDs with decimal points are NH$_3$ branches.}
\tablenotetext{b}{\footnotesize Total integrated intensity of NH$_3$ (1,1) emission.}
\tablenotetext{c}{\footnotesize 500 $\mu$m dust continuum of {\it Herschel Space Observatory}. Typical $rms$ is 0.4 MJy/sr.}
\tablenotetext{d}{\footnotesize Statistical uncertainty of peak column density is only due to uncertainties of the 500 $\mu$m dust continuum flux density and $T_k$. $T_d=T_k$ is assumed. }
\tablenotetext{f}{\footnotesize Branches are omitted and noted as `-'. }
\label{table1}
\end{deluxetable}


\newpage
\begin{deluxetable}{ c c c c c c }
  \tablewidth{0pt}
  \tablecaption{Physical Properties of NH$_3$ sources}
  \startdata
  \hline
ID$^a$	& T$_k^b$ & $\sigma_v^c$ & Mass from dust &  Virial Mass$^d$ & Mass from NH$_3^e$ \\
		& [K] & [km s$^{-1}$] & [M$_\odot$]  &  [M$_\odot$]     &  [M$_\odot$] \\
\hline
1	&	10.24$_{-0.42}^{0.17}$	&	0.199$_{-0.005}^{0.005}$	&	0.18	&	0.45	&	0.36	\\
1.5	&	9.92$_{-0.86}^{0.52}$	&	0.201$_{-0.005}^{0.011}$	&	4.56	&	2.21	&	3.91	\\
2	&	9.66$_{-0.08}^{0.09}$	&	0.197$_{-0.005}^{0.005}$	&	0.06	&	0.19	&	0.09	\\
3	&	10.25$_{-0.84}^{0.93}$	&	0.204$_{-0.008}^{0.014}$	&	4.38	&	1.45	&	1.03	\\
4	&	8.49$_{-0.99}^{0.99}$	&	0.210$_{-0.012}^{0.011}$	&	0.31	&	0.34	&	0.05	\\
5	&	9.65$_{-1.34}^{1.52}$	&	0.252$_{-0.016}^{0.018}$	&	1.74	&	1.93	&	0.71	\\
6	&	9.47$_{-0.50}^{0.78}$	&	0.227$_{-0.020}^{0.005}$	&	0.77	&	1.15	&	0.50	\\
6.5	&	9.49$_{-0.52}^{0.59}$	&	0.201$_{-0.009}^{0.028}$	&	1.85	&	1.61	&	1.39	\\
7	&	9.48$_{-0.55}^{0.43}$	&	0.194$_{-0.005}^{0.006}$	&	0.88	&	0.98	&	0.77	\\
7.5	&	9.29$_{-0.77}^{0.78}$	&	0.202$_{-0.009}^{0.022}$	&	9.52	&	3.89	&	7.58	\\
8	&	9.26$_{-0.21}^{0.16}$	&	0.196$_{-0.003}^{0.002}$	&	0.26	&	0.47	&	0.37	\\
8.5	&	8.99$_{-0.54}^{0.52}$	&	0.198$_{-0.005}^{0.008}$	&	3.20	&	1.88	&	3.64	\\
9	&	8.61$_{-0.24}^{0.30}$	&	0.197$_{-0.005}^{0.007}$	&	1.02	&	0.89	&	1.35	\\
10	&	9.63$_{-2.11}^{2.12}$	&	0.224$_{-0.012}^{0.029}$	&	0.41	&	0.61	&	0.25	\\
11	&	10.00$_{-0.54}^{0.39}$	&	0.203$_{ 0.011}^{0.012}$	&	0.10	&	0.58	&	0.12	\\
12	&	9.29$_{-0.38}^{0.67}$	&	0.213$_{-0.017}^{0.026}$	&	0.95	&	0.98	&	0.98	\\
12.5	&	9.49$_{-0.57}^{0.94}$	&	0.220$_{-0.021}^{0.024}$	&	1.32	&	1.47	&	1.24	\\
13	&	10.90$_{-1.41}^{1.06}$	&	0.236$_{-0.013}^{0.009}$	&	0.04	&	0.31	&	0.03	\\
13.25	&	9.68$_{-0.75}^{1.18}$	&	0.227$_{-0.022}^{0.036}$	&	2.79	&	2.67	&	2.12	\\
14	&	9.68$_{-0.37}^{0.59}$	&	0.226$_{-0.008}^{0.010}$	&	0.23	&	0.61	&	0.13	\\
15	&	8.84$_{-0.50}^{0.39}$	&	0.217$_{-0.014}^{0.035}$	&	0.21	&	0.50	&	0.15	\\
15.5	&	9.50$_{-1.06}^{0.64}$	&	0.206$_{-0.012}^{0.039}$	&	2.73	&	2.02	&	1.88	\\
16	&	9.66$_{-0.30}^{0.27}$	&	0.196$_{-0.003}^{0.007}$	&	0.53	&	0.75	&	0.46	\\
17	&	9.31$_{-0.54}^{0.59}$	&	0.196$_{-0.005}^{0.007}$	&	1.62	&	1.22	&	1.97	\\
18	&	9.75$_{-0.40}^{0.43}$	&	0.222$_{-0.009}^{0.006}$	&	0.12	&	0.40	&	0.09	\\
18.5	&	9.48$_{-0.47}^{0.54}$	&	0.220$_{-0.010}^{0.006}$	&	0.99	&	1.09	&	0.66	\\
19	&	9.11$_{-0.47}^{0.71}$	&	0.214$_{-0.006}^{0.006}$	&	0.07	&	0.22	&	0.05	\\
19.25	&	9.45$_{-1.24}^{0.99}$	&	0.217$_{-0.010}^{0.013}$	&	4.06	&	2.15	&	2.09	\\
20	&	9.50$_{-1.29}^{1.22}$	&	0.217$_{-0.007}^{0.012}$	&	0.30	&	0.66	&	0.15	\\
21	&	9.17$_{-1.30}^{1.36}$	&	0.201$_{-0.016}^{0.016}$	&	1.92	&	1.30	&	0.70	\\
21.925	&	9.80$_{-0.82}^{0.87}$	&	0.203$_{-0.012}^{0.017}$	&	8.97	&	3.82	&	9.74	\\
22	&	9.74$_{-0.75}^{0.58}$	&	0.203$_{-0.011}^{0.014}$	&	1.89	&	1.39	&	1.73	\\
22.85	&	9.85$_{-0.77}^{0.82}$	&	0.203$_{-0.012}^{0.017}$	&	6.84	&	3.55	&	8.97	\\
23	&	10.39$_{-0.54}^{0.51}$	&	0.210$_{-0.011}^{0.005}$	&	0.95	&	1.11	&	1.45	\\
23.75	&	9.91$_{-0.80}^{0.94}$	&	0.203$_{-0.012}^{0.018}$	&	4.77	&	3.04	&	7.16	\\
24	&	9.08$_{-0.16}^{0.15}$	&	0.191$_{-0.003}^{0.003}$	&	0.88	&	0.77	&	2.14	\\
24.5	&	9.20$_{-0.23}^{1.05}$	&	0.192$_{-0.003}^{0.013}$	&	1.81	&	1.38	&	3.63	\\
25	&	10.26$_{-0.78}^{0.48}$	&	0.204$_{-0.009}^{0.009}$	&	0.27	&	0.53	&	0.45	\\
26	&	9.75$_{-1.22}^{1.31}$	&	0.202$_{-0.012}^{0.012}$	&	0.24	&	0.77	&	0.16	\\
27	&	10.27$_{-1.09}^{1.01}$	&	0.203$_{-0.011}^{0.004}$	&	0.08	&	0.45	&	0.05	\\
27.75	&	10.63$_{-1.31}^{0.99}$	&	0.204$_{-0.014}^{0.010}$	&	0.85	&	1.65	&	0.73	\\
28	&	11.32$_{-0.58}^{0.95}$	&	0.213$_{-0.005}^{0.010}$	&	0.15	&	0.70	&	0.15	\\
28.5	&	10.64$_{-1.32}^{0.99}$	&	0.205$_{-0.014}^{0.010}$	&	0.77	&	1.66	&	0.68	\\
29	&	10.03$_{-0.97}^{0.67}$	&	0.195$_{-0.007}^{0.008}$	&	0.19	&	0.68	&	0.18	\\
30	&	9.92$_{-1.18}^{1.08}$	&	0.212$_{-0.005}^{0.008}$	&	1.08	&	1.39	&	0.99	\\
30.5	&	9.77$_{-1.09}^{1.14}$	&	0.220$_{-0.012}^{0.022}$	&	3.09	&	2.56	&	2.03	\\
31	&	9.55$_{-0.52}^{0.49}$	&	0.231$_{-0.011}^{0.014}$	&	0.50	&	0.93	&	0.29	\\
32	&	8.95$_{-1.32}^{1.96}$	&	0.207$_{-0.019}^{0.018}$	&	0.26	&	0.40	&	0.08	\\
33	&	9.54$_{-0.58}^{0.93}$	&	0.206$_{-0.011}^{0.013}$	&	2.49	&	1.87	&	4.61	\\
34	&	13.39$_{-0.91}^{0.57}$	&	0.270$_{-0.015}^{0.027}$	&	0.07	&	0.69	&	0.04	\\
35	&	9.22$_{-1.11}^{0.80}$	&	0.208$_{-0.005}^{0.007}$	&	0.36	&	0.60	&	0.17	\\
36	&	9.81$_{-0.92}^{1.62}$	&	0.205$_{-0.012}^{0.012}$	&	0.19	&	0.57	&	0.11	\\
37	&	9.83$_{-0.62}^{0.50}$	&	0.209$_{-0.008}^{0.018}$	&	1.82	&	1.61	&	2.97	\\
38	&	9.55$_{-1.48}^{0.62}$	&	0.212$_{-0.007}^{0.006}$	&	0.31	&	0.62	&	0.26	\\
39	&	9.55$_{-0.56}^{0.68}$	&	0.199$_{-0.006}^{0.014}$	&	2.89	&	1.51	&	3.22	\\
\enddata
\tablenotetext{a}{\footnotesize IDs with decimal points are branches in dendrograms.}
\tablenotetext{b}{\footnotesize Median T$_k$ of the NH$_3$ leaves and branches. Subscripts and superscripts are 15.9th \%tile and 84.1th \%tile of the T$_k$ distribution within NH$_3$ leaves and branches.}
\tablenotetext{c}{\footnotesize Median $\sigma_v$ of the NH$_3$ leaves and branches with the mean molecular weight $\mu$=2.33. Subscripts and superscripts are 15.9th \%tile and 84.1th \%tile of the $\sigma_v$ distribution within NH$_3$ leaves and branches.}
\tablenotetext{d}{\footnotesize The NH$_3$ leaves and branches are assumed to be homoeoidal ellipsoids for the virial mass calculation.}
\tablenotetext{e}{\footnotesize The fractional abundance of NH$_3$ is assumed to be 1.5 $\times$ 10$^{-9}$ (Typical fractional abundance ranges from 10$^{-9}$ to 10$^{7}$, e.g. Jijina et al. 1999; Tafalla et al. 2004).}
\label{table2}
\end{deluxetable}
\clearpage

\newpage
\begin{deluxetable}{ c c c c c  }
  \tablewidth{0pt}
  \tablecaption{Geometrical Properties of NH$_3$ sources}
  \startdata
  \hline
ID$^a$	&Short axis	&Long axis	    &Axis Ratio & Phase Angle$^b$\\
		& [10$^{-2}$ pc] & [10$^{-2}$ pc]&  & [Degree]  \\
\hline
1	&	0.65	&	2.27	&	0.287	&	38.9	\\
1.5	&	3.79	&	10.1	&	0.376	&	43.9	\\
2	&	0.46	&	0.8	&	0.581	&	44.8	\\
3	&	2.8	&	5.98	&	0.469	&	152.5	\\
4	&	0.91	&	1.61	&	0.568	&	140.1	\\
5	&	2.75	&	5.47	&	0.505	&	34.4	\\
6	&	1.47	&	4.21	&	0.348	&	151.4	\\
6.5	&	1.67	&	9.69	&	0.172	&	1.3	\\
7	&	1.69	&	4.85	&	0.348	&	18.1	\\
7.5	&	3.37	&	27.2	&	0.124	&	24.6	\\
8	&	1.14	&	1.99	&	0.571	&	20.1	\\
8.5	&	2.21	&	11.1	&	0.200	&	23.7	\\
9	&	1.34	&	4.57	&	0.294	&	16.2	\\
10	&	3.00	&	3.69	&	0.813	&	170.3	\\
11	&	0.99	&	1.6	&	0.621	&	4.3	\\
12	&	2.09	&	3.44	&	0.606	&	25.4	\\
12.5	&	2.01	&	5.64	&	0.357	&	37.8	\\
13	&	0.64	&	8.27	&	0.775	&	179.6	\\
13.25	&	3.03	&	11.3	&	0.268	&	38.9	\\
14	&	1.16	&	1.92	&	0.606	&	36.9	\\
15	&	0.87	&	1.82	&	0.476	&	38.0	\\
15.5	&	2.86	&	9.85	&	0.291	&	173.8	\\
16	&	1.27	&	3.67	&	0.345	&	143.9	\\
17	&	4.40	&	4.74	&	0.935	&	173.6	\\
18	&	0.72	&	1.33	&	0.541	&	160.2	\\
18.5	&	2.12	&	3.64	&	0.581	&	43.4	\\
19	&	0.69	&	0.69	&	1.00	&	30.0	\\
19.25	&	5.30	&	6.97	&	0.763	&	23.4	\\
20	&	0.97	&	2.56	&	0.379	&	165.5	\\
21	&	2.67	&	5.73	&	0.467	&	13.1	\\
21.925	&	4.64	&	20.4	&	0.228	&	174.4	\\
22	&	3.81	&	5.1	&	0.752	&	171.7	\\
22.85	&	3.57	&	21.1	&	0.170	&	178.9	\\
23	&	2.11	&	4.19	&	0.505	&	177.0	\\
23.75	&	2.74	&	19.6	&	0.140	&	176.8	\\
24	&	2.25	&	3.26	&	0.694	&	20.45	\\
24.5	&	2.27	&	7.36	&	0.309	&	177.5	\\
25	&	1.13	&	2.09	&	0.543	&	156.7	\\
26	&	0.87	&	4.31	&	0.202	&	4.5	\\
27	&	0.68	&	2.09	&	0.328	&	41.2	\\
27.75	&	2.6	&	7.41	&	0.352	&	20.7	\\
28	&	1.82	&	2.32	&	0.787	&	37.8	\\
28.5	&	2.16	&	8.22	&	0.263	&	16.4	\\
29	&	1.19	&	3.32	&	0.361	&	4.2	\\
30	&	1.94	&	5.96	&	0.326	&	142.3	\\
30.5	&	2.47	&	12	&	0.207	&	147.9	\\
31	&	1.52	&	2.88	&	0.529	&	137.7	\\
32	&	1.19	&	1.38	&	0.870	&	138.8	\\
33	&	3.58	&	7.46	&	0.481	&	7.0	\\
34	&	0.84	&	1.57	&	0.538	&	159.2	\\
35	&	1.29	&	2.26	&	0.571	&	27.3	\\
36	&	1.01	&	2.41	&	0.418	&	28.3	\\
37	&	3.4	&	5.98	&	0.571	&	2.4	\\
38	&	1.28	&	2.23	&	0.571	&	17.5	\\
39	&	4.45	&	5.79	&	0.769	&	146.8	\\
\enddata
\tablenotetext{a}{\footnotesize IDs with decimal points are branches in dendrograms.}
\tablenotetext{b}{\footnotesize The phase angle of NH$_3$ leaves with respect to the Galactic longitude. For the phase angle with respect to the equatorial plane. 62.6$^\circ$ should be added.}
\label{table3}
\end{deluxetable}

\clearpage

\newpage

\newpage
\begin{figure*}
\includegraphics[angle=0,scale=0.8]{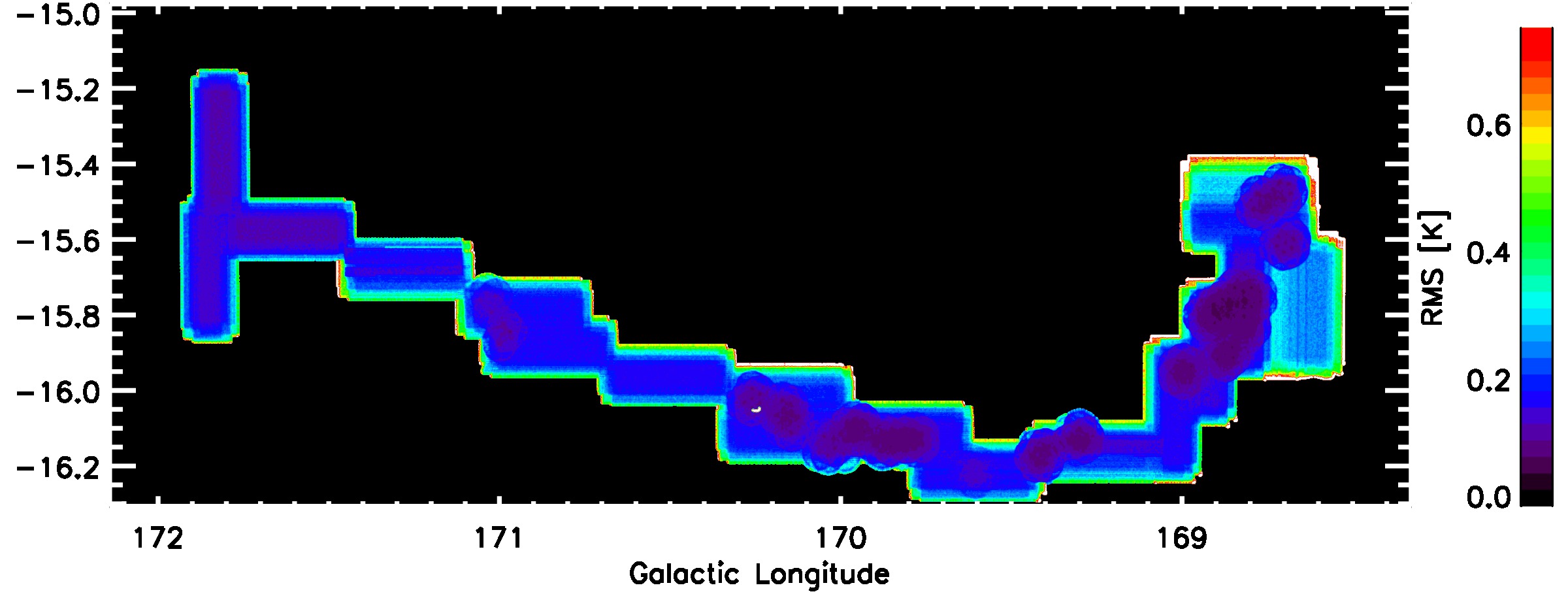}
\caption{$rms$ of the ammonia $T_{\rm mb}$ map at a spectral resolution of 6.1 kHz (corresponding to 0.077 km s$^{-1}$), obtained by smoothing two spectral channels.}
\label{fig2}
\end{figure*}

\newpage
\begin{figure*}
\includegraphics[angle=90,scale=0.7]{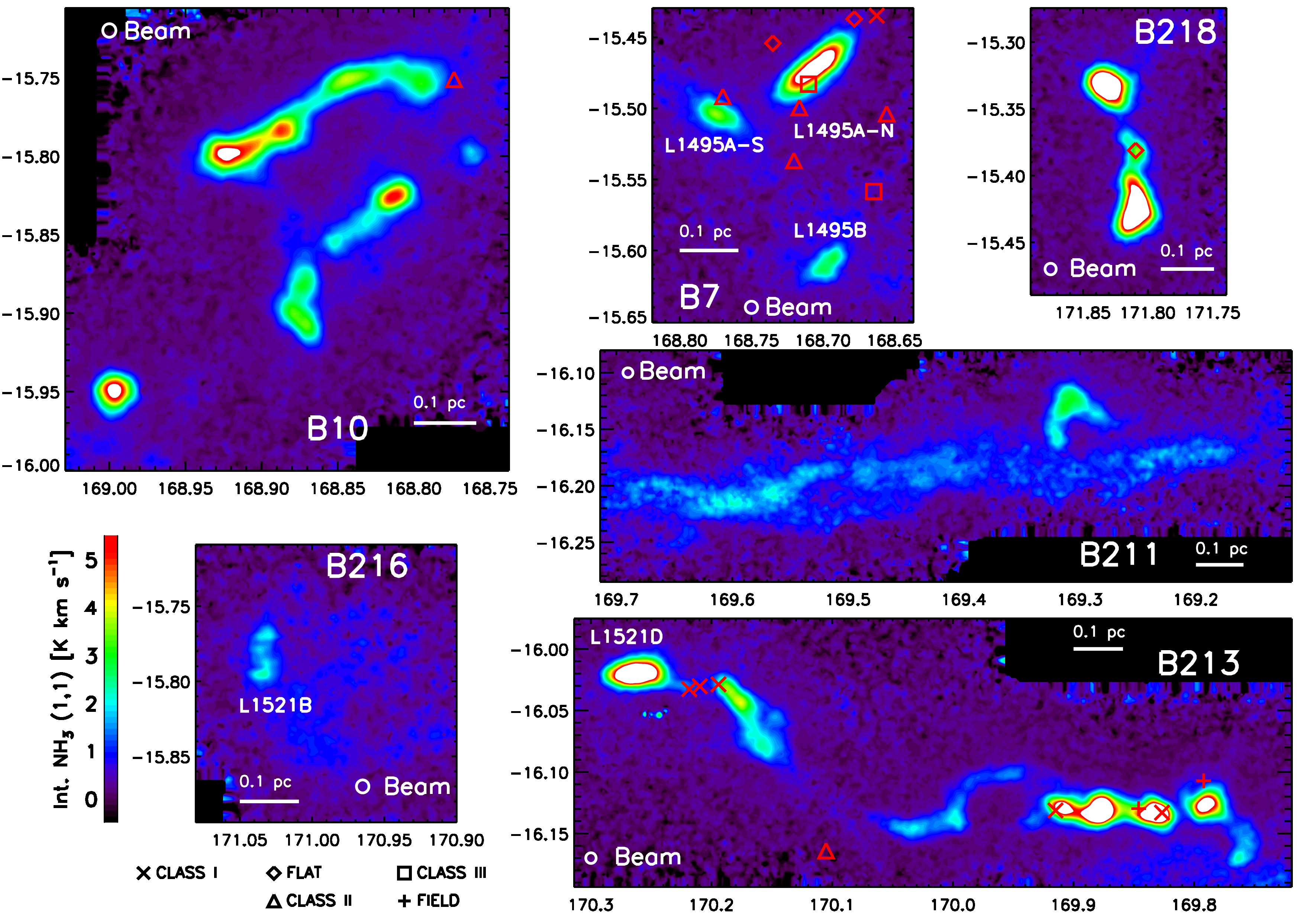}
\caption{{Maps of integrated NH$_3$ (1,1) emission and positions of known protostars. X-axis is galactic longitude and Y-axis is galactic latitude. The names of previously studied cores are indicated in the image.}}
\label{fig5}
\end{figure*}

\newpage
\begin{figure*}
\includegraphics[angle=90,scale=0.7]{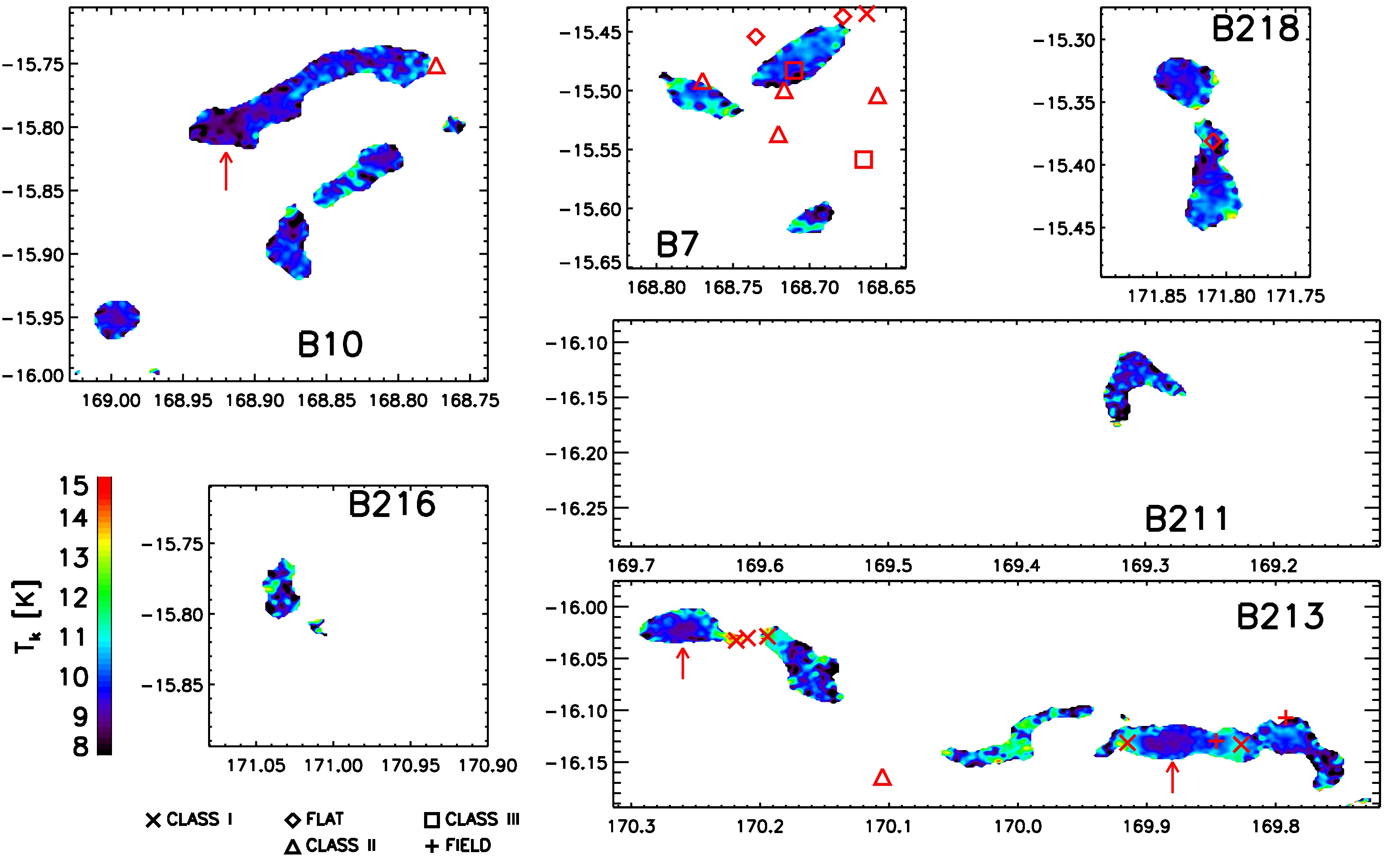}
\caption{{Map of the gas kinetic temperature in L1495-B218 filaments. X-axis is galactic longitude and Y-axis is galactic latitude. Locations of protostars are marked with various symbols depending on the spectral classification of protostars. Red arrows indicate dense cores having gas kinetic temperatures decreasing toward core centers. Regions with multiple velocity components in B211 are suppressed.}}
\label{fig7}
\end{figure*}

\newpage
\begin{figure*}
\includegraphics[angle=90,scale=0.7]{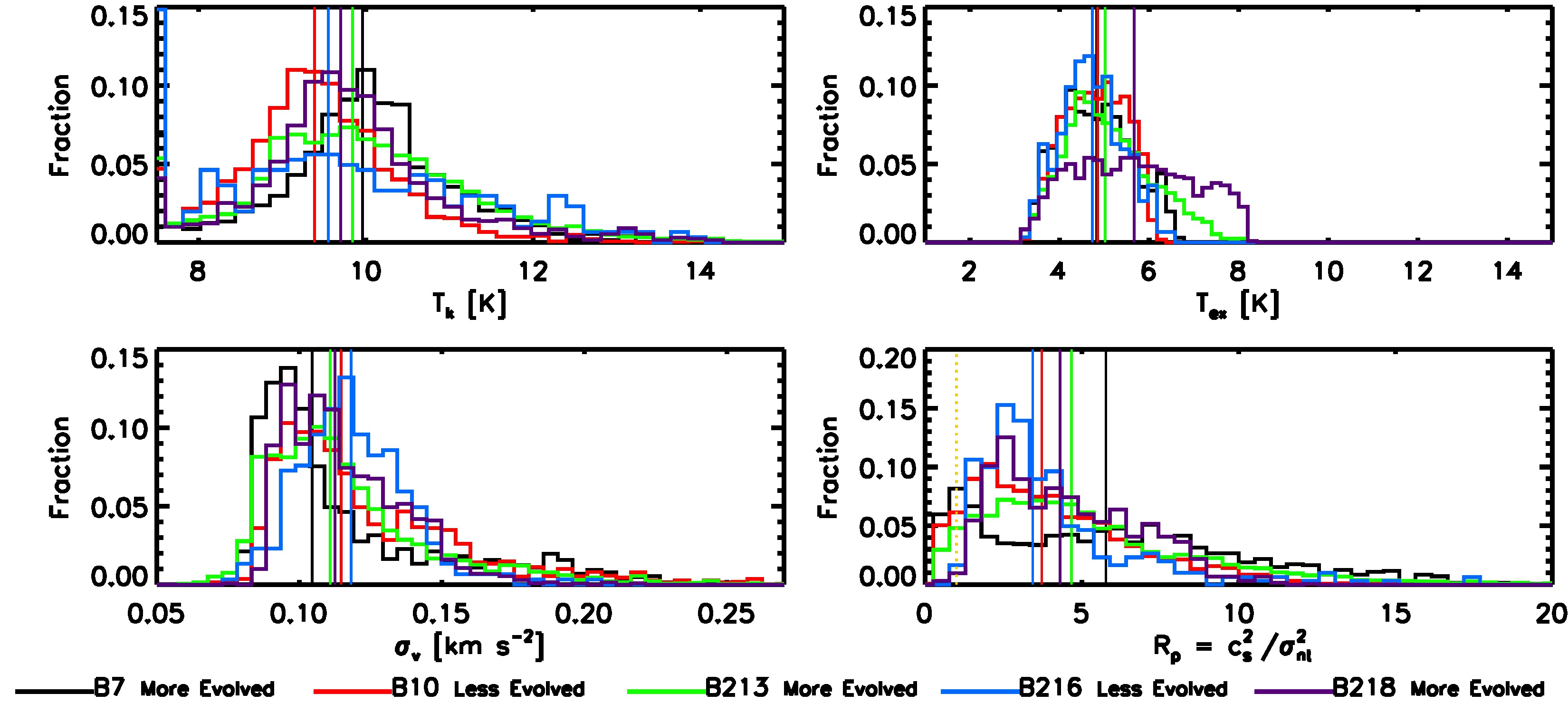}
\caption{{Histograms of physical properties for each region. Vertical lines denote the median values. High fractions at the very left edge of the first panel is due to low signal-to-noise level at the edge of map. Those are excluded in estimating median values. B211 is also excluded due to the presence of multiple velocity components. The dotted vertical line in the fourth panel denote $R_p$ = 1.}}
\label{histograms}
\end{figure*}

\newpage
\begin{figure*}
\includegraphics[angle=0,scale=0.8]{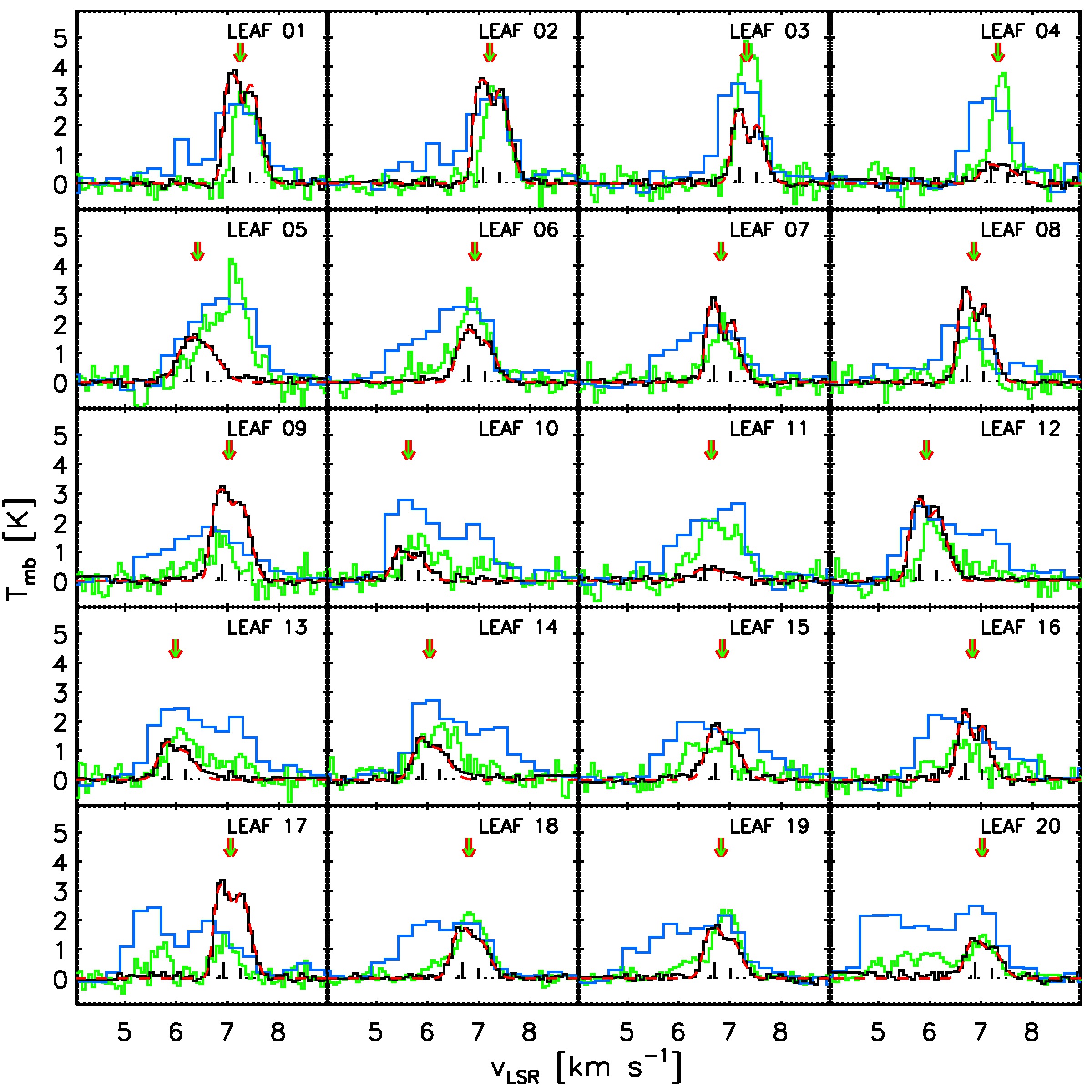}
\caption{{NH$_3$ (1,1) lines (black histograms), the best-fit single component NH$_3$ (1,1) model (dashed red lines), C$^{18}$O 1$-$0 (green histograms; Hacar et al. 2013), and $^{13}$CO 1$-$0 (Goldsmith et al. 2008; Narayanan et al. 2008) at the NH$_3$ peak positions of NH$_3$ leaves identified by CSAR. Black vertical lines denote locations and relative strength of NH$_3$ (1,1) hyperfine lines. Red-green arrows denote the LSR velocities of NH$_3$ leaves. $^{13}$CO and C$^{18}$O have multiple velocity components in some of NH$_3$ leaves but NH$_3$ seems to be confined to one velocity component and is well fitted with a NH$_3$ line model with a single velocity component.}}
\label{vlsr1}
\end{figure*}

\newpage
\begin{figure*}
\includegraphics[angle=0,scale=0.8]{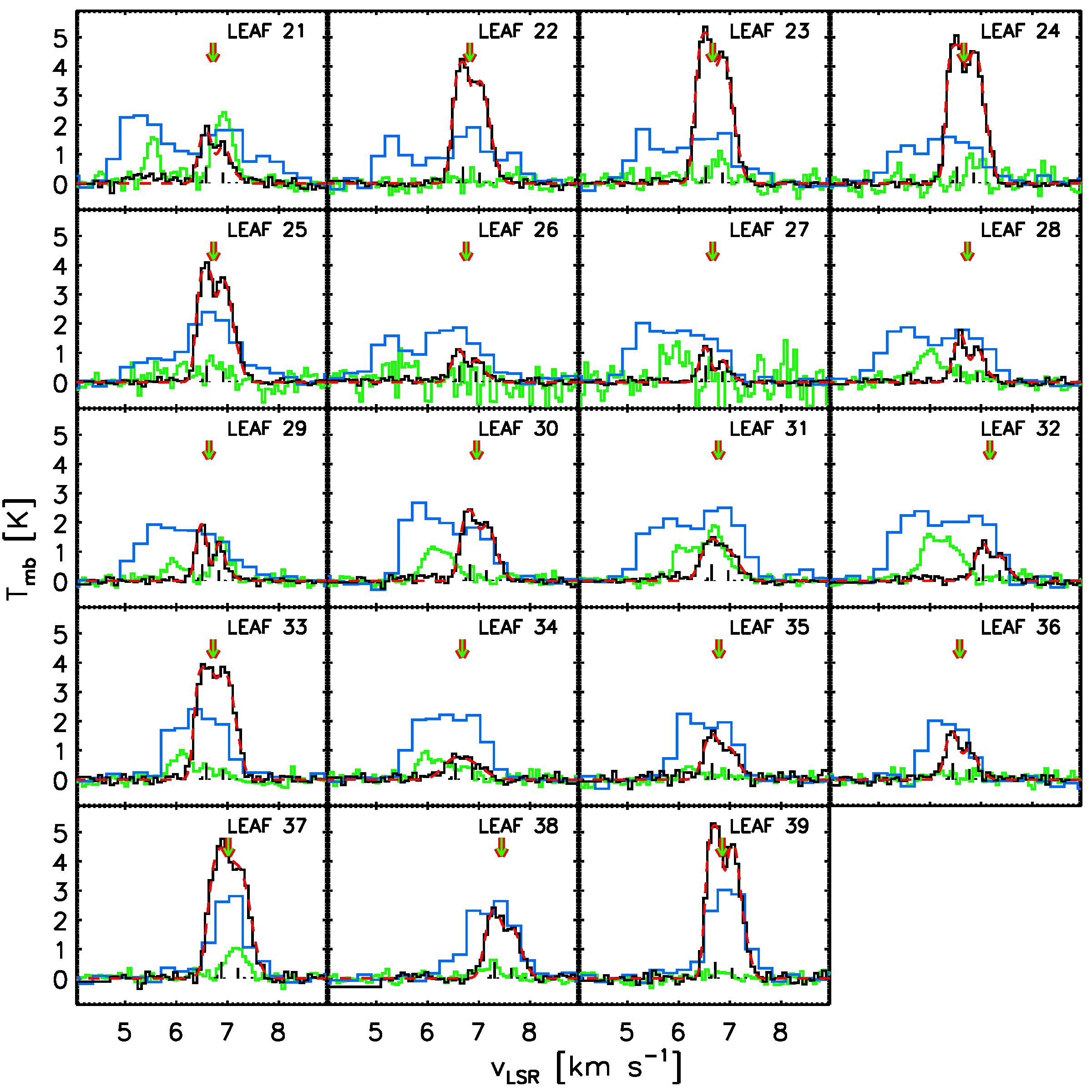}
\caption{{Continued from Fig. \ref{vlsr1}.}}
\label{vlsr2}
\end{figure*}

\newpage
\begin{figure*}
\includegraphics[angle=90,scale=0.7]{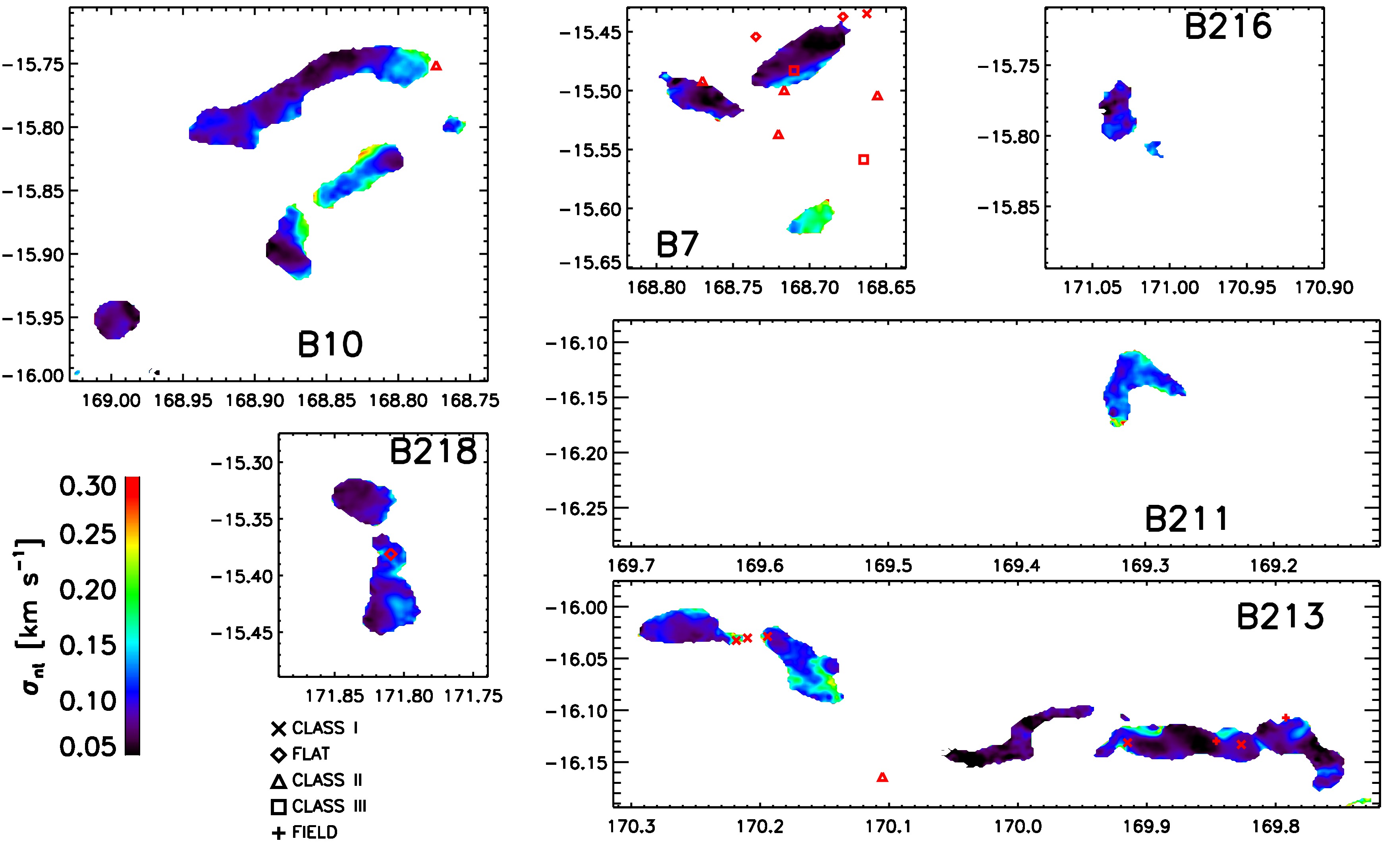}
\caption{{Velocity dispersions of nonthermal components. X-axis is the galactic longitude and Y-axis is the galactic latitude. Locations of protostars are marked with various symbols depending on spectral classification of protostars. Regions with multiple velocity components in B211 are suppressed.}}
\label{sigmav}
\end{figure*}

\newpage
\begin{figure*}
\includegraphics[angle=90,scale=0.7]{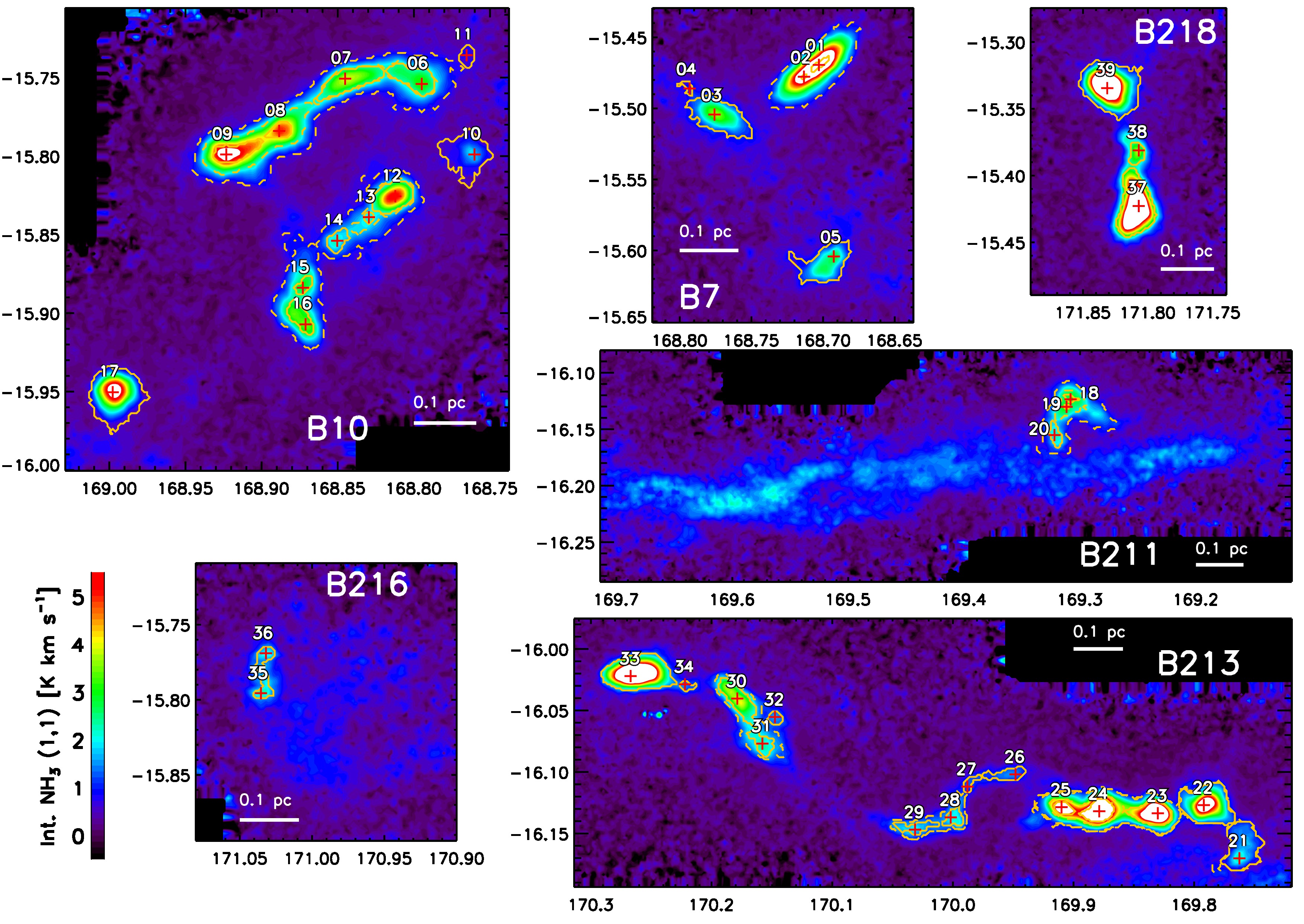}
\caption{{Location of NH$_3$ leaves. Solid and dashed orange contours are NH$_3$ leaves and branches, respectively, identified by CSAR.}}
\label{csar}
\end{figure*}

\newpage
\begin{figure*}
\includegraphics[angle=0,scale=0.9]{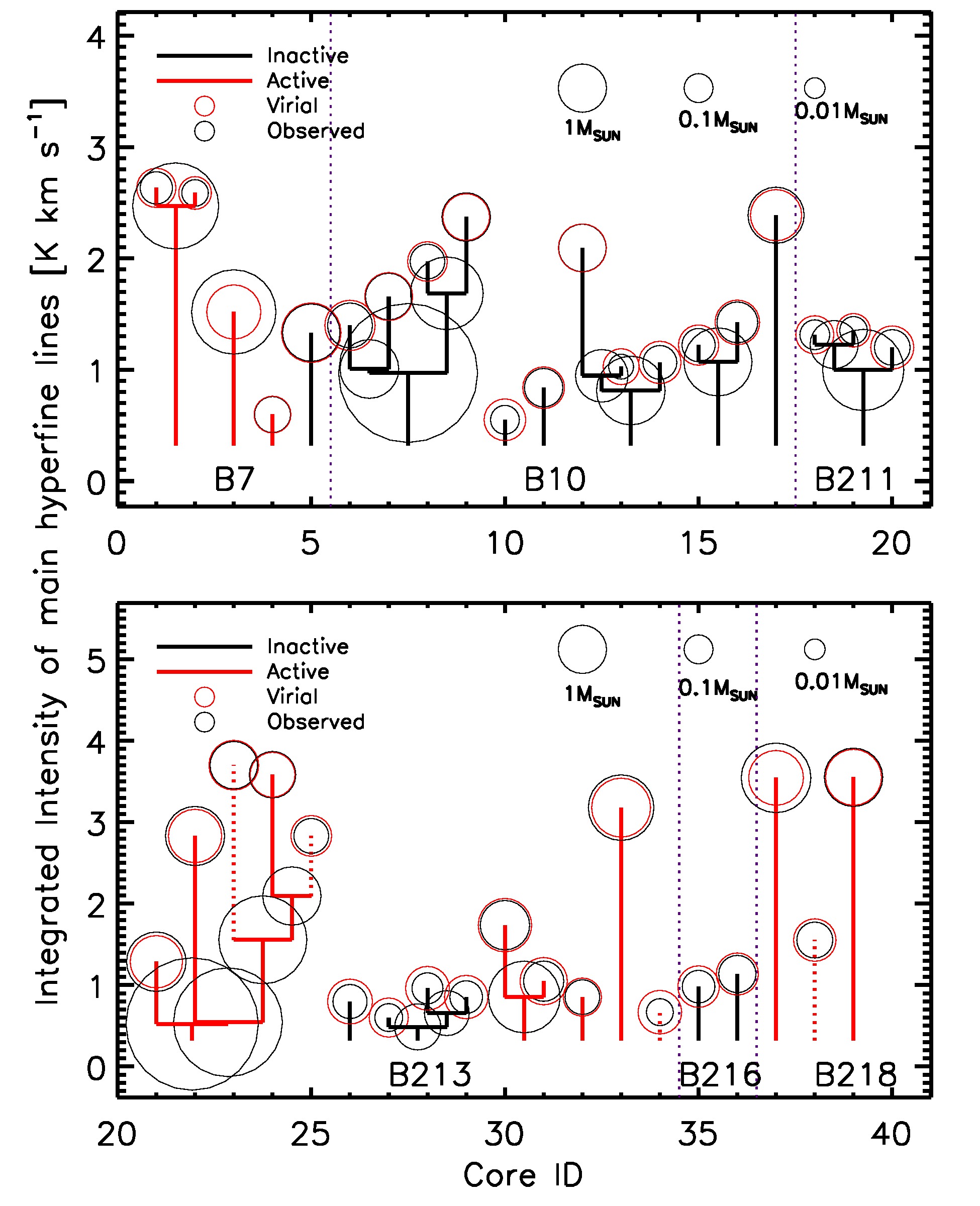}
\caption{{Dendrogram of NH$_3$ sources. Red solid lines denote active leaves (leaves that are embedding a ClassI/Flat spectra protostar within their lowest-level branches or closely associated with a ClassI/Flat spectra protostar within three times their leaf radiuses) and dashed red lines denote active leaves with an embedded protostar. Black circles indicate mass of NH$_3$ sources estimated from 500$\mu m$ dust continuum. Red circles denote the virial mass. Purple dashed vertical lines indicates different subregions.}}
\label{dendro}
\end{figure*}

\newpage
\begin{figure*}
\includegraphics[angle=90,scale=0.7]{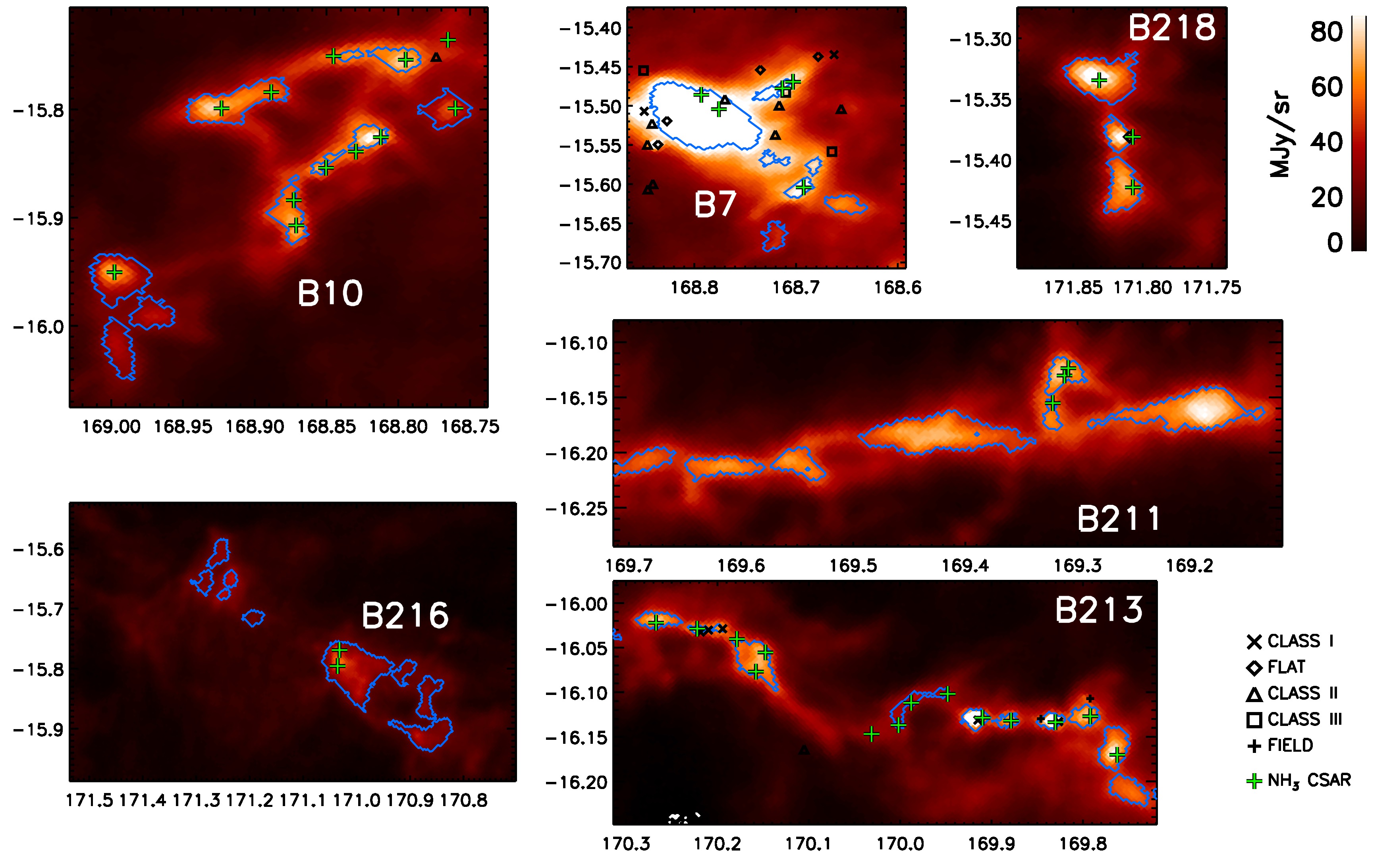}
\caption{{Location of dust and NH$_3$ leaves. Blue contours are boundaries of dust leaves and green-black crosses are the peak positions of NH$_3$ leaves.}}
\label{cores}
\end{figure*}

\newpage
\begin{figure*}
\includegraphics[angle=0,scale=0.8]{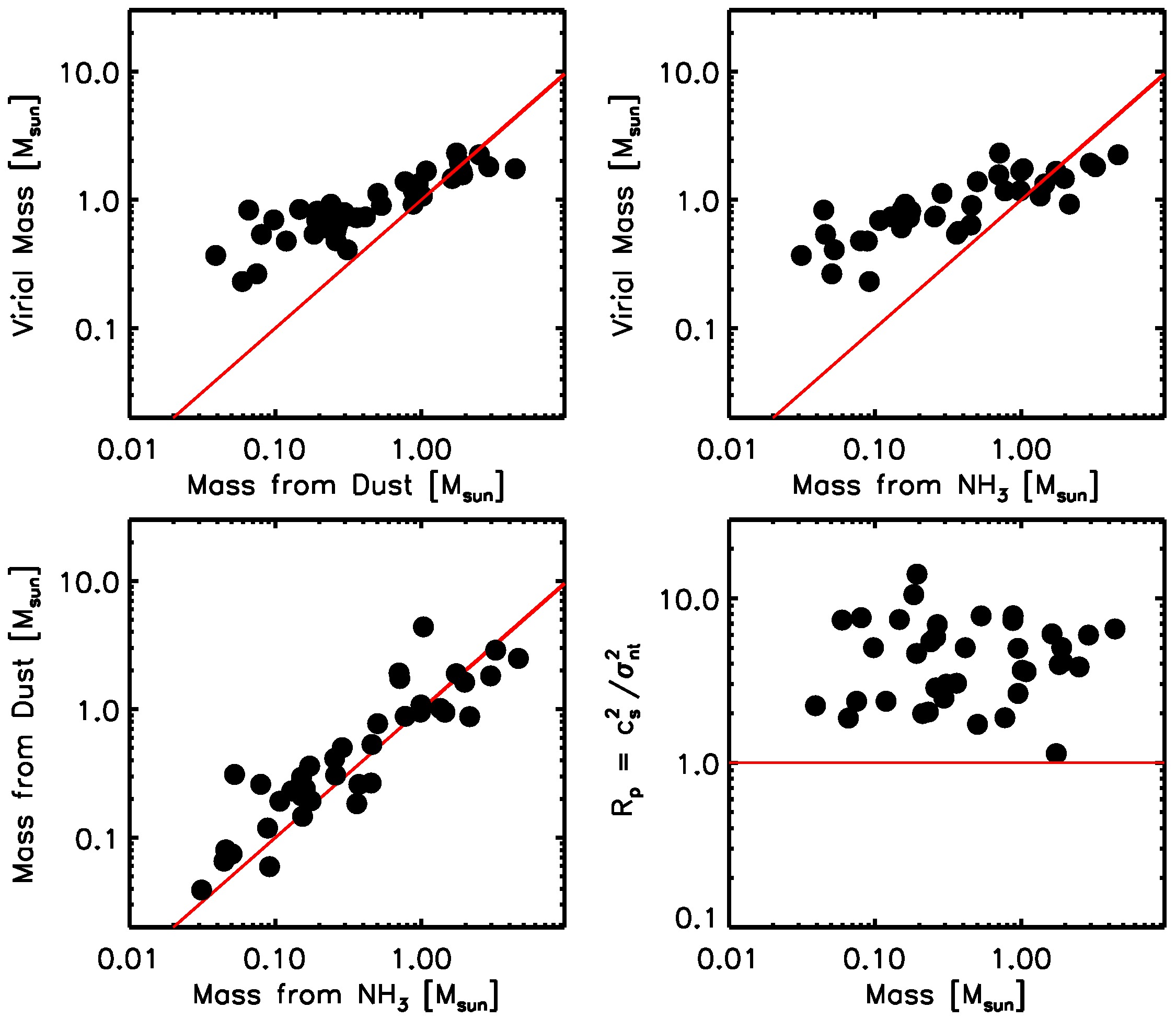}
\caption{{Relations between masses (the first, second, and third panels). The red line denotes $x$=$y$. Ratio of thermal support to nonthermal support (the fourth panel). The red line denotes $R_p$ = 1.}}
\label{cmass}
\end{figure*}

\newpage
\begin{figure*}
\includegraphics[angle=0,scale=0.8]{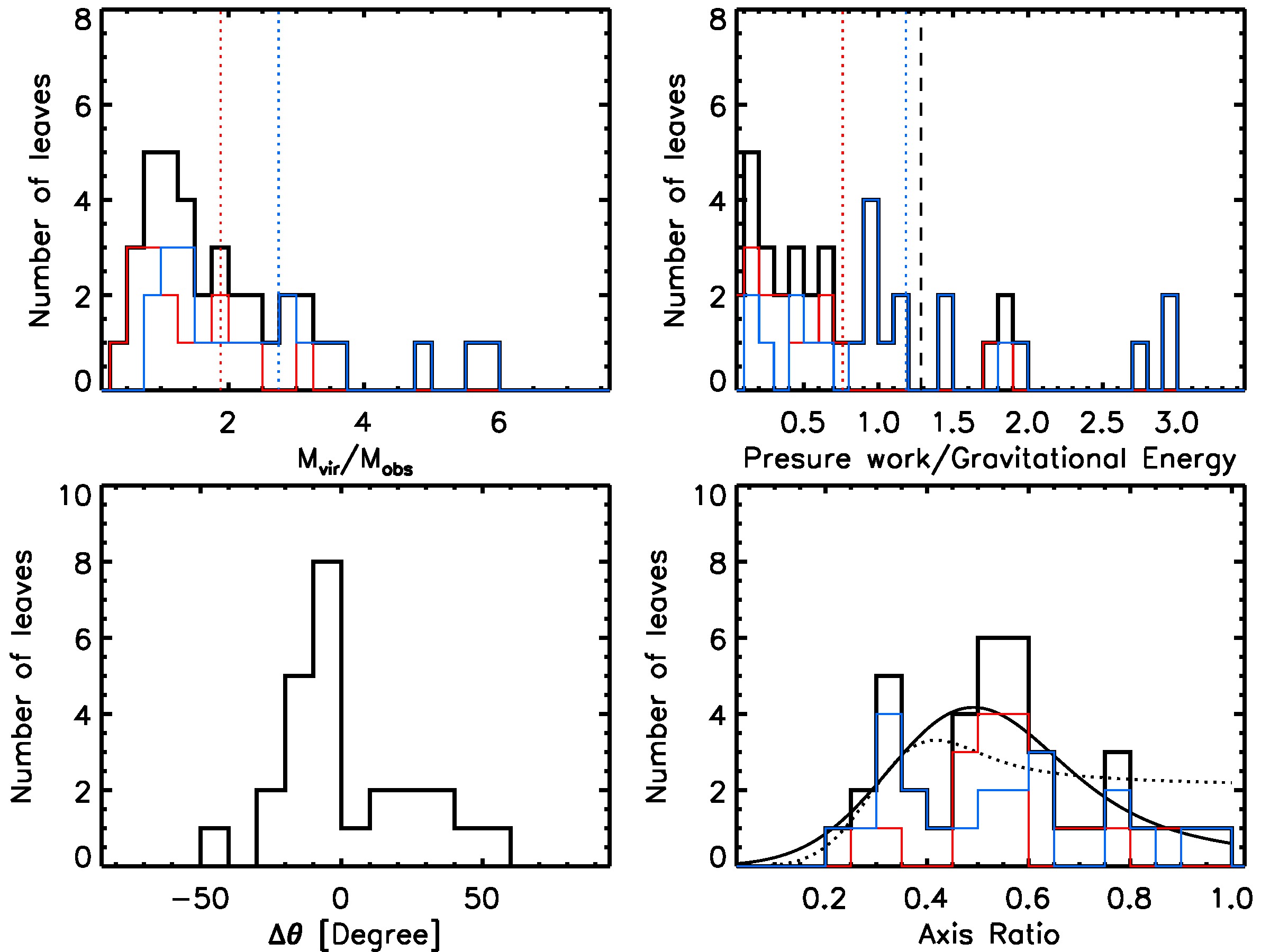}
\caption{{Distribution of properties of the NH$_3$ leaves. Red histograms denote active leaves, and blue histograms are inactive leaves. The first panel is the ratio of the virial mass to the observed mass. The red and blue vertical lines are the mean ratios of active and inactive leaves, respectively. The second panel is the ratio of the work done by pressure to the gravitational energy at the surface of NH$_3$ leaves. The red and blue vertical lines are the mean ratios of active and inactive leaves, respectively. The black dashed vertical line is the energy ratio of the critical Bonnor-Ebert sphere. The third panel is the angle difference of the NH$_3$ leaves to their branches. The fourth panel is the axis ratio of the NH$_3$ leaves deduced from the principal axis analysis. The solid and dashed lines are the best-fit probability density of apparent axis ratio of prolate and oblate spheroids with a uniform orientation angle distribution within [-80$^\circ$, 80$^\circ$], respectively.}}
\label{chist}
\end{figure*}

\newpage
\begin{figure*}
\includegraphics[angle=0,scale=0.8]{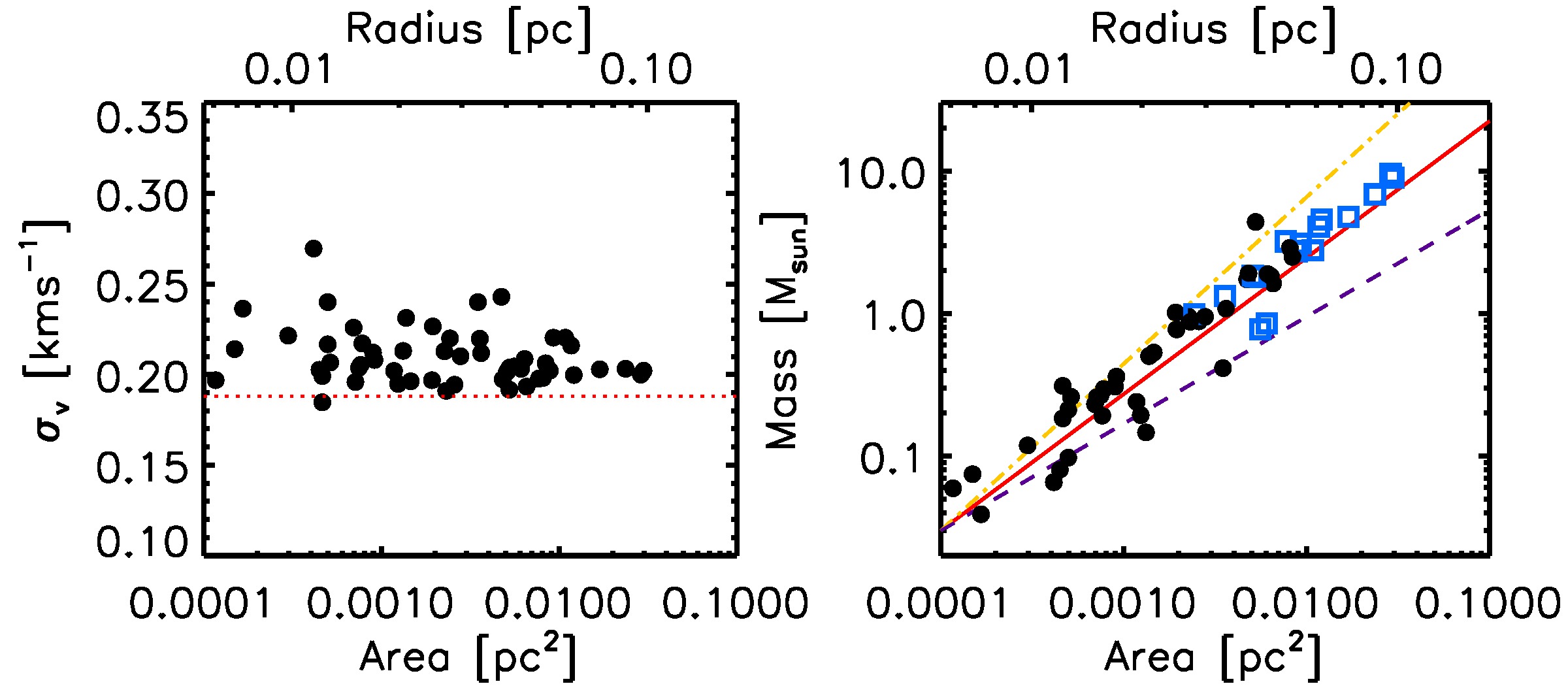}
\caption{{Relations between physical properties of NH$_3$ sources. The left panel shows the relation of the velocity dispersion ($\mu$=2.33) and the size of the NH$_3$ leaves. The red dotted line denotes the thermal velocity dispersion when T$_k$ = 10 and $\mu$ = 2.33. The right panel is the mass-size relation of the NH$_3$ sources. Black filled circles are the NH$_3$ leaves and blue boxes are the NH$_3$ branches. The red solid line is our best fit of M$\sim r^{1.9}$ to the NH$_3$ sources. The purple dashed line is M$\sim r^{1.5}$ from Kauffman et al. (2010) and the orange dashed-dotted line is M$\sim r^{2.35}$ from Kirk et al. (2013).}}
\label{crel}
\end{figure*}

\newpage
\begin{figure*}
\includegraphics[angle=0,scale=0.8]{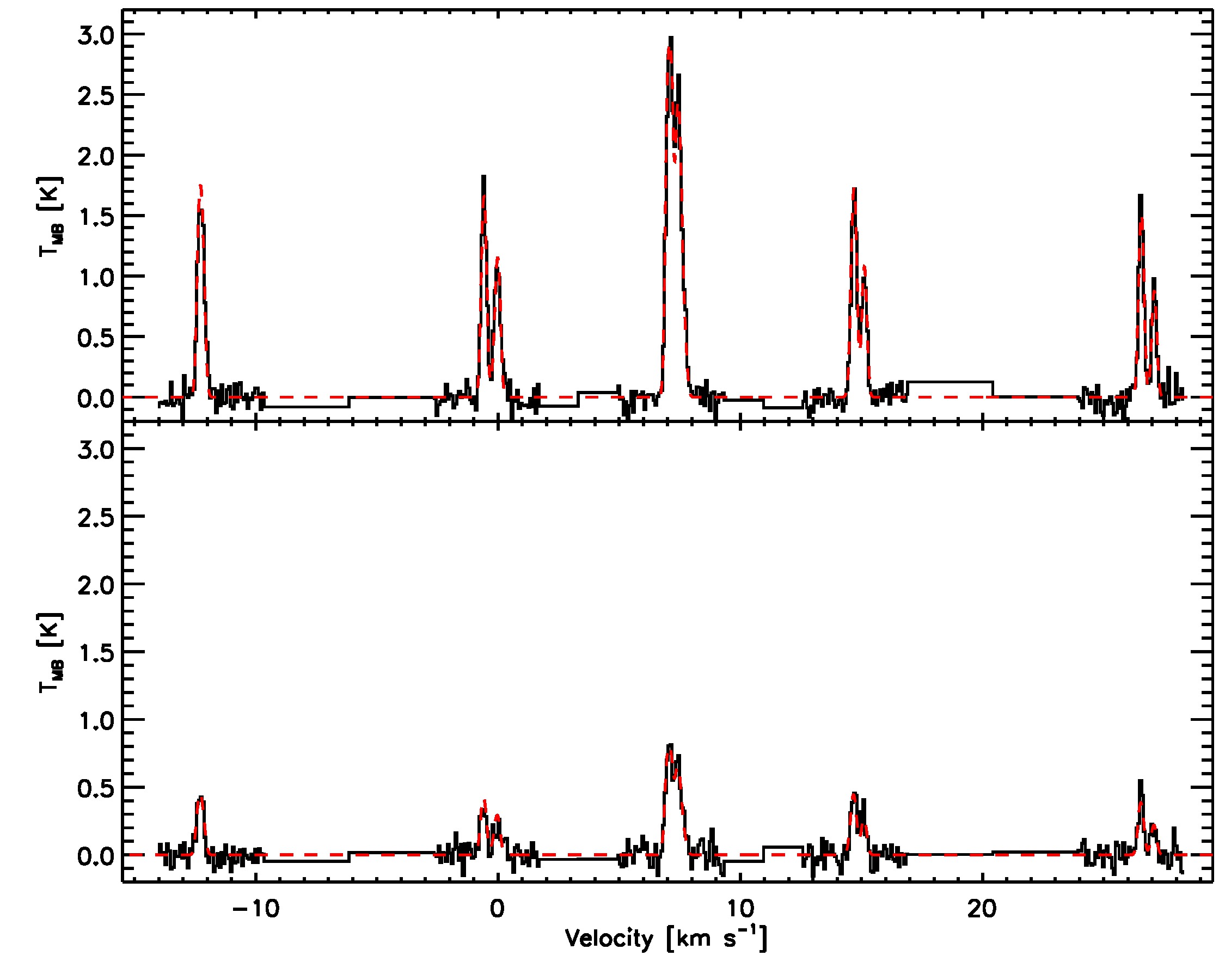}
\caption{{Examples of spectra model fitting (red dashed lines) to observed lines (black solid lines). The top panel shows NH$_3$ (1,1) lines having a peak signal to noise ratio (SNR) equal to 32, and the bottom panel shows NH$_3$ (1,1) lines having a peak signal to noise ratio equal to 7.5.}}
\label{line_examples}
\end{figure*}

\newpage
\begin{figure*}
\includegraphics[angle=0,scale=0.6]{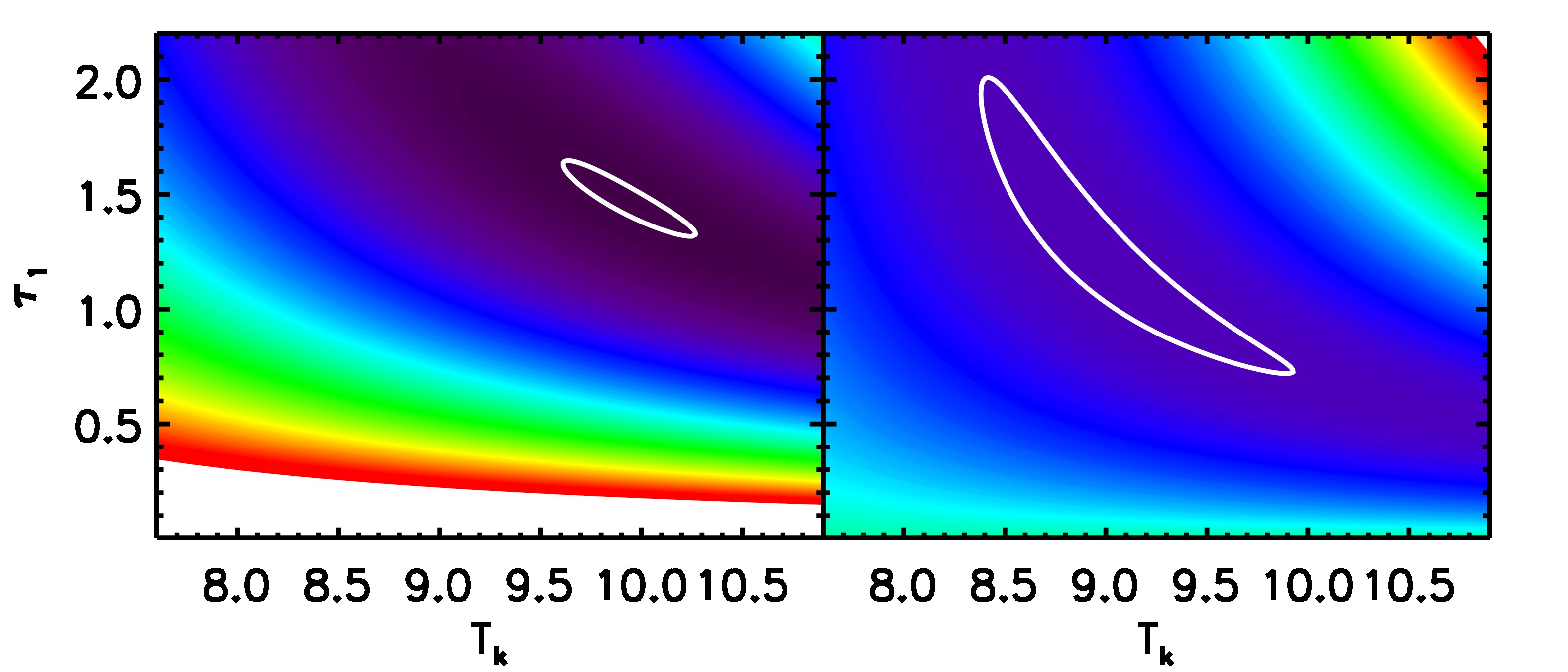}
\caption{{Examples of reduced the $\chi^2$ space in T$_k$ $vs.$ $\tau_1$. The left panel shows the $\chi^2$ space of fitting the line with a peak SNR of 32, and the right panel shows $\chi^2$ space of fitting the line with a peak SNR of 7.5. The white contours are 1-$\sigma$ uncertainty limits.}}
\label{chisq_examples}
\end{figure*}

\newpage
\begin{figure*}
\includegraphics[angle=0,scale=0.7]{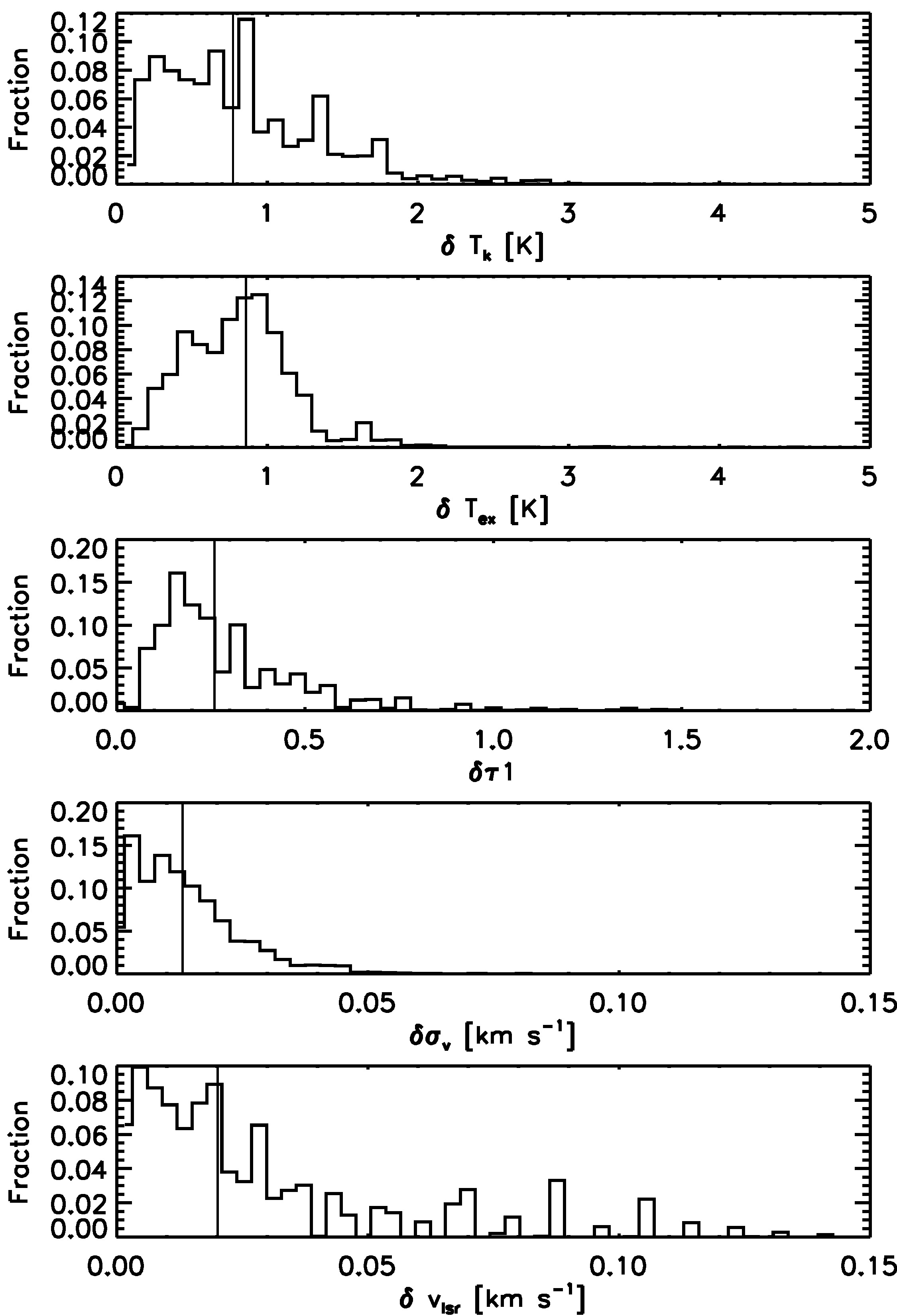}
\caption{{Distributions of 1-$\sigma$ uncertainties in the spectra fittings. Vertical lines represent the median values.}}
\label{errors}
\end{figure*}

\end{document}